\documentclass[10pt]{article}

\usepackage{amsmath}
\usepackage{graphicx}
\usepackage{natbib}
\usepackage{url} 
\usepackage{enumerate}
\usepackage{amsfonts,amssymb,mathrsfs,dsfont,amsthm}
\usepackage{color}
\usepackage{bbm,bm}
\usepackage{comment}
\usepackage{bigints}
\usepackage[normalem]{ulem}
\usepackage[utf8]{inputenc}
\usepackage{float}
\usepackage{multirow}
\usepackage{array}
\RequirePackage[colorlinks,citecolor=blue,urlcolor=blue]{hyperref}
\usepackage{chngcntr}

\newtheorem{theorem}{Theorem}
\newtheorem{lemma}{Lemma}[section]
\theoremstyle{definition}
\newtheorem{remark}{Remark}[section]

\def\l{\left}
\def\r{\right}

\newcolumntype{C}[1]{>{\centering\arraybackslash}p{#1}}

\newcommand{\ddr}{\mathrm{d}}

\newcommand{\fine}{\hfill $\Box$}
\newcommand{\comillas}[1]{``\,#1\,''}
\definecolor{iblue}{rgb}{0.1,0,0.75}
\definecolor{ired}{rgb}{0.9,0,0.1}
\definecolor{gray}{rgb}{0.5, 0.5, 0.5}

\newcommand{\Ind}{\mathds{1}}

\newcommand{\tp}{\tilde{p}}
\newcommand{\tx}{\tilde{x}}

\newcommand{\Esp}{\mathbb{E}}
\newcommand{\Prob}{\mathbb{P}}
\newcommand{\X}{\mathbb{X}}

\newcommand{\N}{{\mathbb{N}}}
\newcommand{\B}{\mathcal{B}}
\newcommand{\D}{\mathcal{D}}

\newcommand{\Be}{\mathsf{Be}}
\newcommand{\Dir}{\mathsf{Dir}}
\newcommand{\pr}{\mathbbm{p}}

\newcommand{\clA}{\mathcal{A}}

\newcommand{\rest}{\cdots}
\newcommand{\oal}{\overline{\alpha}}

\allowdisplaybreaks

\title{\bf Gibbs sampling for mixtures in order of appearance: the ordered allocation sampler}

\author{
  Pierpaolo De Blasi\\
  Collegio Carlo Alberto and ESOMAS Department,\\ University of Torino, Italy\\
  \texttt{pierpaolo.deblasi@unito.it} \\
  \and
  Mar\'ia F. Gil--Leyva\\
  Department of Probability and Statistics, IIMAS--UNAM, M\'exico\\
  \texttt{marifer@sigma.iimas.unam.mx} \\
}

\date{}

\begin{document}
\maketitle

\begin{abstract}
Gibbs sampling methods are standard tools to perform posterior inference for mixture models. These have been broadly classified into two categories: marginal and conditional methods. 
While conditional samplers are more widely applicable than marginal ones, they may suffer from slow mixing in infinite mixtures, where some form of truncation, either deterministic or random, is required. In mixtures with random number of components, the exploration of parameter spaces of different dimensions can also be challenging. We tackle these issues by expressing the mixture components in the random order of appearance in an exchangeable sequence directed by the mixing distribution. We derive a sampler that is straightforward to implement for mixing distributions with tractable size-biased ordered weights, and that can be readily adapted to mixture models for which marginal samplers are not available. In infinite mixtures, no form of truncation is necessary. As for finite mixtures with random dimension, a simple updating of the number of components is obtained by a blocking argument, thus, easing challenges found in trans-dimensional moves via Metropolis-Hastings steps. Additionally, sampling occurs in the space of ordered partitions with blocks labelled in the least element order, which endows the sampler with good mixing properties. The performance of the proposed algorithm is evaluated in a simulation study.
\end{abstract}

\noindent%
\vfill

\newpage

\section{Introduction}\label{sec:1}

Mixture models represent one of the most successful applications of Bayesian methods. 
Bayesian inference proceeds by placing a prior on the mixing distribution, whose atoms and their sizes represent the mixture component parameters and the weights, respectively. An important issue is that the number of components is rarely known in advance. The nonparametric approach consists in modelling the mixing distribution with infinitely many support points, and infer the number of components through the number of groups observed in the data \citep[e.g.][]{Esc:Wes:95}. 
Another alternative is to assign a prior to this unknown quantity \citep[cf.][]{Ric:Gre:97,Mil:Har:18}.
Posterior inference for mixture models is customarily based on Gibbs sampling methods which have been broadly classified into two categories: {\it marginal} and {\it conditional} samplers. 
Marginal methods \citep{Esc:Wes:95,Nea:00} are termed this way because they partially integrate out the mixing distribution and exploit the generalized P\'olya urn scheme representation \citep{Bla:Mac:73,Pit:06} of the prediction rule of a sample from the mixing distribution. By doing so marginal samplers avoid dealing with a potentially infinite or random model dimension. They are well suited for models such as the Dirichlet \citep{Fer:73,Esc:Wes:95}, the Pitman-Yor \citep{Pit:Yor:97,Ish:Jam:01} and mixtures of finite mixtures \citep{Mil:Har:18}. However, they are challenging to adapt to mixing priors without a tractable prediction rule. 
In alternative, one can use conditional methods which include the mixing distribution and update it as a component of the sampler. While being more widely applicable they bring some issues in their design and implementation.
In infinite mixture models, some sort of truncation either deterministic or random is necessary to avoid dealing with infinitely many mixture components. Finite dimensional approximations of the mixing prior were proposed in \citep{Ish:Jam:01}. Of random truncation type 
are the exact conditional samplers derived by \cite{Walk:07,Kall:etal:11} and \cite{Pap:Rob:08}. It has been observed that they require Metropolis-Hastings steps that swap components' labels so to speed up mixing  \citep{Port:etal:06,Pap:Rob:08}. As for mixtures with a random number of components, the main challenge of conditional samplers is the need to explore parameter spaces of different dimensions. The standard method is the reversible jump MCMC algorithm \citep{Ric:Gre:97} but this can be difficult to implement.

In this paper we contribute both methodologically and computationally to mitigate these issues by developing a novel conditional Gibbs sampling method named the \textit{ordered allocation sampler}. The sampler works with the mixture components  in the random order in which they are discovered.
To derive it we use in depth the theory of species sampling models set forth in \citet{Pit:95,Pit:96,Pit:96ims} where, in particular, it is established that the law of the weights in order of appearance corresponds to the distribution of the weights that is invariant under size-biased permutations. This one admits a simple stick-breaking representation for the Dirichlet and the Pitman-Yor processes. 
Working with this specific rearrangement of mixture components allows us to exploit a (conditional) prediction rule in the sampler. Thus, it bears similarities with marginal methods such as the fact that mixing takes place in the space of partitions and not in the space of cluster's labels as it occurs in other conditional samplers \citep{Port:etal:06}. Empirical studies confirm that this endows our sampler with nice mixing properties.
A second major advantage of our proposal is that, since $n$ data points can not be generated from more than $n$ distinct components, at most the sampler needs to update the first $n$ components in order of appearance. 
This is especially relevant for infinite mixture models as it avoids truncation. In particular, for Pitman-Yor processes with slowly decaying weights, our sampler proves to be very convenient computationally wise. 
A third important consequence is that marginalization over the 
weights in order of appearance yields exactly the {\it exchangeable partition probability function} \citep[EPPF][]{Pit:96ims}. For mixtures with random dimension, this translates to a simple way of updating the number of components without resorting to a reversible jump step. 
Finally, as other conditional methods, the ordered allocation sampler allows direct inference on the mixing distribution and can be adapted to a wide range of mixing priors.

The rest of the paper is organized as follows. In Section \ref{sec:2} we provide background theory on species sampling priors in mixture models. It will set the stage for 
the ordered allocation sampler. In Section \ref{sec:3} we first derive the sampler for models with tractable size-biased permuted weights, and later we show how to adapt it when the law of this arrangement of the weights is not available in explicit form. In Section \ref{sec:4} we illustrate the performance of our sampler with well-known real and simulated datasets. Some concluding remarks and discussion points are brought in Section \ref{sec:5}. Proofs, technical details and additional illustrations are in the Appendix.


\section{Species sampling models}\label{sec:2}


In Bayesian mixture models we model exchangeable data, $(y_i) = (y_i)_{i=1}^n$, taking values in a Borel space, $(\mathbb{Y},\B(\mathbb{Y}))$, as conditionally independent and identically distributed (iid) from
\begin{equation}\label{eq:mix_mod}
Q(y) = \int_{\X}g(y\mid x)P(\ddr x) = \sum_{j=1}^m p_j\, g(y\mid x_j),
\end{equation}
where $g(\cdot\mid x)$ is a density for each $x$, and the mixing distribution, $P = \sum_{j=1}^m p_j\delta_{x_j}$, is an almost surely discrete random probability measure over the Borel parameter space $(\X,\B(\X))$. {\it Species sampling models}, introduced and studied by \citet{Pit:96ims}, constitute a very general class of random probability measures that provide a convenient prior specification for the mixing distribution $P$. 
In species sampling models, $P = \sum_{j=1}^m p_j\delta_{x_j}$, the atoms, $(x_j) = (x_1,\ldots,x_m)$, are iid from a diffuse distribution, $\nu$, over $(\X,\B(\X))$, and are independent of $(m,(p_j))$. The weights $(p_j) = (p_1,\ldots,p_m)$ are positive random variables with $\sum_{j=1}^m p_j = 1$ almost surely, and the number of support points, $m$, can be finite, infinite or random.
A sequence, $(\theta_i) = (\theta_i)_{i=1}^{\infty}$, is a \emph{species sampling sequence} driven by $P$ if it is exchangeable and the almost sure limit of the empirical distributions, 
  $P = \lim_{n \to \infty}n^{-1}\sum_{i=1}^n \delta_{\theta_i},$
is a species sampling model. 
By de Finetti's theorem, the latter is equivalent to the existence of a species sampling model, $P$, 
such that given $P$, $(\theta_i)$ are conditionally iid according to $P$ \citep[cf. Theorem 1.1 and Proposition 1.4 of][]{Kall:05}. 
The laws of $(\theta_i)$ and $P$ determine each other, and both are fully determined by that of $(m,(p_j))$, and the diffuse distribution, $\nu$. A key aspect to note is that the law of $P$ is invariant under weights permutations. That is, $P = \sum_{j=1}^m p_j \delta_{x_j}$ is equal in distribution to $\sum_{j=1}^m p_{\rho(j)} \delta_{x_j}$ for every permutation $\rho$ of $\{1,\ldots,m\}$, which means that working with an ordering of the weights or another does not change the mixing prior. This is reflected through the so called exchangeable partition probability function (EPPF), given by
\begin{equation}\label{eq:eppf_sum}
  \pi(n_1,\ldots,n_k) 
  = \sum_{(j_1,\ldots,j_k)} \Esp
  \bigg(\prod_{i=1}^k p_{j_i}^{n_i}\bigg),
\end{equation}
where the sum ranges over all $k$-tupples 
of distinct positive integers, and
$p_j = 0$ for $j > m$. In fact, $\pi(n_1,\ldots,n_k)$ describes the probability that a sample,  $(\theta_1,\ldots,\theta_n)$, of size $n = \sum_{j=1}^k n_j$, exhibits exactly $k$ distinct values, with corresponding frequencies, $n_1,\ldots,n_k$  \citep{Pit:96ims,Pit:06}. Whenever $\pi(n_1,\ldots,n_k)$ can be computed in closed form, the prediction rule, $\Prob[\theta_{i+1}\in \cdot \mid \theta_1,\ldots,\theta_i]$, of $(\theta_i)$ is available and can be described in terms of a generalized P\'olya urn scheme \citep[cf][]{Bla:Mac:73}.

The invariance under permutations of $\sum_{j=1}^m p_{\rho(j)} \delta_{x_j}$, as well as the complexity of computing the unordered sum in \eqref{eq:eppf_sum}, have motivated the study of weights permutations that simplify the analysis. 
An ordering of the weights of paramount importance is the {\it size-biased permutation}, $(\tp_j) = (\tp_1,\ldots,\tp_m)$, given by $\tp_j = p_{\alpha_j}$, and $(\alpha_j) = (\alpha_1,\ldots,\alpha_m)$ defined by
\begin{equation}\label{eq:size-biased_pick}
\begin{split}
  \Prob[\alpha_1 = j\mid (p_j)] 
  &= p_j, \\
  \Prob[\alpha_l = j\mid (p_j),\alpha_1,\ldots,\alpha_{l-1}]
  &=\frac{p_j}{1-\sum_{i=1}^{l-1}p_{\alpha_i}}
  \Ind_{\{j \not\in \{\alpha_1,\ldots,\alpha_{l-1}\}\}}, 
  \quad 2\leq l\leq m.
\end{split}
\end{equation}
In other words, $(\alpha_j)$ and $(\tp_j)$ are sampled without replacement from $\{1,\ldots,m\}$ and $(p_j)$, respectively, with probabilities $(p_j)$. By construction the distribution of $(\tp_j)$ is \emph{invariant under size-biased permutations}.
As shown by \cite{Pit:95,Pit:96}, the EPPF \eqref{eq:eppf_sum} can be computed through
\begin{equation}\label{eq:eppf_sb}
  \pi(n_1,\ldots,n_k) 
  = \Esp\bigg[\prod_{j=1}^k \tp_j^{n_j-1}
  \prod_{j=1}^{k-1}\bigg(1-\sum_{l=1}^j \tp_l\bigg)\bigg],
\end{equation}
hence, if the distribution of $(\tp_j)$ is available, it becomes easier to compute $\pi(n_1,\ldots,n_k)$. 
Another advantage of working with size-biased permutations is that $\tp_j$ coincides with the long-run proportion of indexes $i$ such that $\theta_i = \tx_j$, where $\tx_j$ is the $j$th distinct value to appear in $(\theta_i)$. Furthermore, the conditional law of $(\theta_i)$ given $(\tx_j) = (\tx_1,\ldots,\tx_m)$ and $(\tp_j)$ admits a simple prediction rule as detailed next.

\begin{theorem}\label{theo:rep_sss}
Let $P = \sum_{j=1}^m p_j \delta_{x_j}$ be a species sampling model over the Borel space $(\X,\B(\X))$ and let $(\theta_i) = (\theta_i)_{i=1}^\infty$ be a sequence with values in $(\X,\B(\X))$. Define the $j$th distinct value to appear in $(\theta_i)$ through $\tx_j = \theta_{M_j}$, where
$M_{j} = \min\{i > M_{j-1} : 
  \theta_i \not\in \{\tx_1,\ldots,\tx_{j-1}\}\}$,
for $j \geq 2$ and $M_1 = 1$. Then $(\theta_i)$ is an species sampling sequence driven by $P$ if and only if the following hold:
\begin{enumerate}[i.]
\item
$(\theta_i)$ exhibits $m$ distinct values, $(\tx_j) = (\tx_1,\ldots,\tx_m)$, in order of appearance, and $(\tx_j)$ are iid from $\nu$. Furthermore, for $j \leq m$, $\tx_j = x_{\alpha_j}$ where $(\alpha_j)$ satisfies \eqref{eq:size-biased_pick}.
\item
The almost sure limits
  $$
  \tilde{p}_j 
  = \lim_{n \to \infty}\frac{|\{i \leq n: \theta_i 
  = \tilde{x}_j\}|}{n}, \quad j \leq m,
  $$
exist, $\tp_j > 0$, $\sum_{j=1}^m \tp_j = 1$ almost surely, and $(\tp_j) = (\tp_1,\ldots,\tp_m)$ is invariant under size-biased permutations. Moreover, $(\tp_j)$ is given by $\tp_j = p_{\alpha_j}$, with $(\alpha_j)$ as in 2.i. 
\item
$\theta_1 = \tx_1$, and the conditional prediction rule of $(\theta_i)$ given $(\tp_j)$ and $(\tx_j)$ is
  $$
  \Prob[\theta_{i+1}\in \cdot
  \mid  (\tp_j),(\tx_j),\theta_1,\ldots,\theta_i]
  = \sum_{j=1}^{k_i}\tilde{p}_j\delta_{\tilde{x}_j} 
  + \bigg(1- \sum_{j=1}^{k_i}\tilde{p}_j\bigg)
  \delta_{\tilde{x}_{k_i +1}},
  $$
for every $i \geq 1$, where $k_i$ is the number of distinct values in $(\theta_1,\ldots,\theta_i)$.
\item
$m$, $(\tp_j)$, $(p_j)$ and $(\alpha_j)$ are independent of elements in $(\tx_j)$.
\end{enumerate}
\end{theorem}

Theorem \ref{theo:rep_sss} is based on theory laid down in \cite{Pit:95,Pit:96ims}, nonetheless, we provide a self-contained proof  
in Appendix \ref{sec:supp_proof}, due to the crucial role it plays in the derivation of the new sampler.


The canonical example of species sampling models in Bayesian nonparametric statistics is the Dirichlet process \citep{Fer:73}. It has $m=\infty$ support points
and its size-biased permuted weights, $(\tp_j)$, admit the stick-breaking representation
\begin{equation}\label{eq:sb}
  \tp_1 = v_1, \quad 
  \tp_j = v_j\prod_{i=1}^{j-1}(1-v_i), \quad j \geq 2,
\end{equation}
where $(v_j) = (v_j)_{j=1}^{\infty}$ are iid from the Beta distribution $\Be(1,\theta)$ \citep{Seth:94}. 
The Dirichlet model can be generalized to the two-parameter $(\sigma,\theta)$-model \citep{Pit:06} which features size-biased permuted weights as in \eqref{eq:sb} with independent $v_j \sim \Be(1-\sigma,\theta+j\sigma)$ according to one of the following two regimes:
\begin{enumerate}[a)]
\item 
$\sigma \in [0,1)$ and $\theta > -\sigma$. In this case $m = \infty$ and the species sampling model $P$ has been named the Pitman-Yor process by \citet{Ish:Jam:01} after \citet{Pit:Yor:97}. Evidently the choice $\sigma = 0$ reduces to a Dirichlet process. 
\item
Given $m \in \mathbb{N}$, $\sigma = -\gamma < 0$, and $\theta = m\gamma$. In agreement with the notation we have established $m$ stands for the number of support points of $P$. It turns out that the law of $(\tp_j)$ corresponds to that of the size-biased permutation of symmetric Dirichlet weights, $(p_1,\ldots,p_m) \sim \Dir(\gamma,\ldots,\gamma)$ \citep{Pit:96}. When $\gamma$ is fixed and $m$ is random $P$ belongs to the class of Gibbs-type priors \citep[see][for a recent review]{Deb:etal:15}, while the allied mixture model corresponds to the mixture of finite mixtures of \citet{Mil:Har:18}. 
\end{enumerate}

Another type of species sampling models for which a stick-breaking characterization of size-biased weights is available are homogeneous normalized random measures with independent increments \citep[cf.][]{Reg:etal:03,Fav:etal:16}. 
Unfortunately, such characterization remains elusive for most species sampling models used in mixture modelling, examples are finite dimensional approximations of the Pitman-Yor process \citep{Ish:Jam:01}, the Geometric process \citep{Fue:etal:10}, the probit stick-breaking process \citep{Rod:Dun:11} and exchangeable stick-breaking processes studied by \cite{Gil:Men:21}. For all these species sampling priors the weights can be defined in terms of  a stick-breaking decomposition, $p_j=v_j\prod_{l=1}^{j-1}(1-v_l)$, for some sequence of random variables $(v_j)$ with values in $[0,1]$, yet $(p_j)$ is not invariant under size-biased permutations.

\section{The ordered allocation sampler}\label{sec:3}


As mentioned in Section \ref{sec:2}, in mixture models data points $(y_i) = (y_i)_{i=1}^n$ are treated as conditionally iid from a random density $Q$ as in \eqref{eq:mix_mod}. 
Whenever the mixing distribution, $P$, is a species sampling model we can equivalently assume $y_i\mid \theta_i \sim g(\cdot\mid \theta_i)$, independently for $i \leq n$, where $(\theta_i)$ is a species sampling sequence driven by $P$. In this setting,  marginal samplers integrate out $P$ and exploit the exchangeability of 
$(\theta_i)$ as well as the prediction rule, $\Prob[\theta_{i+1}\in \cdot \mid \theta_1,\ldots,\theta_i]$,  to derive an algorithm for posterior inference \citep[cf.][]{Nea:00,Fav:Teh:13,Mil:Har:18}. Instead, conditional samplers \citep[e.g.][]{Ish:Jam:01,Pap:Rob:08,Kall:etal:11} include the mixing distribution, $P$, and update its atoms, $(x_j)$, and weights, $(p_j)$, as components of the sampler. 
The ordered allocation sampler is a conditional sampler as it includes the mixing distribution, $P$, however similarly to marginal samplers it relies on a prediction rule for species sampling sequences. Explicitly, motivated by Theorem \ref{theo:rep_sss} we work with the atoms, $(\tx_j)$, and weights, $(\tp_j)$, of $P$ in the order in which they were discovered by $(\theta_i)$. As commonly done in other samplers, we augment the model with latent allocation variables that identify each observation, $y_i$, with the mixture component it was sampled from. Here, in accordance with the order of appearance we introduce what we call {\it ordered allocation variables}, $(d_i) = (d_i)_{i=1}^{n}$, given by $d_i = j$ if and only if $y_i$ was sampled from $g(\cdot\mid \tx_j)$, i.e. $\theta_i = \tx_j$. Thus, $\theta_i = \tx_{d_i}$, and
  $y_i \mid ((\tx_j),d_i) \sim g(\cdot\mid \tx_{d_i})$, 
independently for $i \leq n$.
If $k_i$ denotes the number of distinct values in $(\theta_1,\ldots,\theta_i)$ then $k_i$ coincides with $\max\{d_1,\ldots,d_i\}$, and $d_{i+1}$ necessarily takes a value in $\{1,\ldots,k_i+1\}$. More precisely, \emph{iii} of Theorem \ref{theo:rep_sss} allows us to compute 
\begin{equation}\label{eq:pred_OAV}
  d_1 = 1,\quad
  d_{i+1}\mid (\tp_j),d_1,\ldots,d_i
  \sim \sum_{j=1}^{k_i}\tp_j \delta_j 
  + \l(1-\sum_{j=1}^{k_i}\tp_j\r)\delta_{k_i +1},
\end{equation}
for $i \leq n$, independently of elements in $(\tx_j)$ (see also \emph{iv} of Theorem \ref{theo:rep_sss}). This yields the augmented likelihood
\begin{equation}\label{eq:like_sb}
\pr[(y_i),(d_i) \mid (\tp_j),(\tx_j)]
= \prod_{j=1}^{k_n}\tp_j^{n_j-1}
  \bigg(1-\sum_{l=1}^{j-1}\tp_l\bigg)
  \prod_{i \in D_j}g(y_i\mid\tx_{j})
  \Ind_{\mathcal{D}},
\end{equation}
where $k_n = \max\{d_1,\ldots,d_n\}$, $D_j = \{i \leq n: d_i = j\}$, $n_j = |D_j|$, and $\mathcal{D}$ is the event that $\{D_1,\ldots,D_{k_n}\}$ is a partition of $\{1,\ldots,n\}$ with blocks in the \emph{least element order}, in particular $D_j \neq \emptyset$, for $j \leq k_n$, and $\min\l(D_1\r)< \min\l(D_2\r) < \cdots < \min\big(D_{{k_n}}\big)$. The full conditional distributions required at each iteration of the sampler are proportional to the product of \eqref{eq:like_sb} times the prior distributions, $\pr[(\tx_j)]$ and $\pr[(\tp_j)]$, of the atoms and weights of $P$ in order of appearance. We first derive the ordered allocation sampler for those species sampling mixing distributions where the prior of $(\tp_j)$ can be modelled directly as is the case of the $(\sigma,\theta)$-model. Latter we explain how to adapt the sampler for the more general case where the law of $(\tp_j)$ is not available.

\subsection{Ordered allocation sampler for size-biased weights}\label{sec:OASsb}

\subsubsection*{Updating of the ordered allocation variables $(d_i)$:}
\begin{equation}\label{eq:d_post_sb_0}
  \pr[\,d_i = d\mid \rest\,]
  \propto \tp_d\, g(y_i\mid\tx_d) \times \prod_{j=1}^{k_n}
  \tp_j^{-1}\bigg(1-\sum_{l=1}^{j-1}\tp_l\bigg)\Ind_{\D}.
\end{equation}
This is the fundamentally novel part of the algorithm. Differently from other conditional samplers, 
the allocation variables $(d_i)$ can not be updated independently of each other for two main reasons: (i) $k_n = \max\{d_1,\ldots,d_n\}$ might change as a consequence of an update in $d_i$, and (ii) the least element order of $D_1,\ldots,D_{k_n}$ must be preserved, as specified by the indicator $\Ind_{\D}$. Instead, the updating of $(d_i)$ resembles the way marginal algorithms update allocation variables \citep[cf.][]{Nea:00} in the sense that we will update one $d_i$ at a time by conditioning on the current value of the remaining ordered allocation variables. To do so, we first identify the set, $\D_i$, of {\it admissible moves} for $d_i$, which contains all positive integers $d$ for which the event $\D$ remains true after setting $d_i = d$. That is, for $d \in \N$, define $k_n^{(d)} = \max\{k_{-i},d\}$, where $k_{-i} = \max\{d_l:l \neq i\}$,
and add $d \in \D_i$ if, under the assumption $d_i = d$, the sets $D_j= \{l\leq n:d_l = j\}$ are non-empty, for $j \leq k_n^{(d)}$, and satisfy $\min\l(D_1\r)< \min\l(D_2\r) < \cdots < \min\big(D_{{k^{(d)}_n}}\big)$. 
An example that illustrates how to determine $\D_i$ is available in Appendix \ref{sec:supp_example_di}. 
With this notation at hand we can rewrite \eqref{eq:d_post_sb_0}:
\begin{equation}\label{eq:d_post_sb_1}
  \pr[\,d_i = d\mid \rest\,]
  \propto \tp_d\, g(y_i\mid\tx_d) \times 
  \prod_{j=1}^{k_n^{(d)}}\tp_j^{-1}
  \bigg(1-\sum_{l=1}^{j-1}\tp_l\bigg)\Ind_{\{d \in \D_i\}}.
\end{equation}
Next, we need to weight the admissible moves according to \eqref{eq:d_post_sb_1}. To this aim, note that for each $d \in \D_i$, either $k_n^{(d)} = k_{-i}$ or $k_n^{(d)} = k_{-i}+1$. This means that we can divide \eqref{eq:d_post_sb_1} by $\prod_{j=1}^{k_{-i}}\tp_j^{-1}(1-\sum_{l=1}^{j-1}\tp_l)$ and obtain
\begin{equation}\label{eq:d_post_sb}
\pr[\,d_i = d\mid \rest\,] \propto \begin{cases}
\tp_d g(y_i\mid \tx_d) & \text{ if }  k_n^{(d)} = k_{-i},\\
\l(1-\sum_{l=1}^{k_{-i}}\tp_l\r)g(y_i\mid \tx_d) & \text{ if } k_n^{(d)} = k_{-i}+1,
\end{cases}
\end{equation}
for $d \in \D_i$, and $\pr[\,d_i = d\mid \rest\,] = 0$, for $d \not\in \D_i$. Once we have identified $\D_i$, sampling from \eqref{eq:d_post_sb} is straight-forward, as its support is $\D_i \subset \{1,\ldots,n\}$. 

\subsubsection*{Updating of component parameters in order of appearance $(\tx_j)$:}\label{sec:3.1}
\begin{equation}\label{eq:x_post_sb}
\pr[\,\tx_j\mid\rest\,] \propto \prod_{i \in D_j} g(y_i\mid \tx_j)\nu(\tx_j).
\end{equation}
Since $D_j = \emptyset$, for $j > k_n$, we simply  sample $\tx_j \sim \nu$ from its prior distribution. For $j \leq k_n$, sampling from \eqref{eq:x_post_sb} is easy if $g$ and $\nu$ form a conjugate pair. Otherwise, the problem of updating the non-empty components parameters is identical as in conditional samplers and some marginal ones. 
The advantage with respect to conditional algorithms is that the occupied component parameters are precisely the first $k_n$ in the sequence $(\tx_j)$. 

\subsubsection*{Updating of size-biased weights $(\tp_j)$:}\label{sec:3.1}
\begin{equation}\label{eq:p_post_sb}
  \pr[\,(\tp_j)\mid\rest\,] 
  \propto \prod_{j=1}^{k_n}\tp_j^{n_j-1}\l(1-\sum_{l=1}^{j-1}\tp_l\r)
  \,\times\,\pr[(\tp_j)].
\end{equation}
If the stick-breaking representation \eqref{eq:sb} is available, we can update $(\tp_j)$ via sampling $(v_j)$ from its full conditional. Noting that $1-\sum_{l=1}^k\tp_l = \prod_{l=1}^k(1-v_l)$ for each $k \geq 1$, we find
\begin{equation*}
  \pr[\,(v_j)\mid\rest\,]
  \propto \bigg[\prod_{j=1}^{k_n}v_j^{n_j-1}(1-v_j)^{\sum_{l > j}n_l}
  \bigg]\times\pr[(v_j)].
\end{equation*}
For example, for the $(\sigma,\theta)$-model we know that apriori $v_j \sim \Be(1-\sigma,\theta+j\sigma)$, independently, according to one of the two regimes (a) or (b) spelled out in Section \ref{sec:2}. In this case, we update $v_j \sim \Be(n_j - \sigma,\theta + j\sigma+ \sum_{l>j}n_l)$, independently for $j \leq k_n$, and we sample $v_j \sim \Be(1-\sigma,\theta+j\sigma)$, for $j > k_n$, as apriori.

Before we move on, there are two points worth highlighting concerning the updating of $(\tp_j)$. The first one is that while the stick-breaking decomposition simplifies this step it is not a requirement, what is needed is a posterior characterization of the weights in order of appearance. The Pitman-Yor multinomial process studied by \citet{Lij:Pru:Rig:20} illustrates this point. The second crucial remark is that sampling $\tp_j$ and $\tx_j$, for $j>k_n$, is needed for only a few $j$ as required by the occupation of new components when updating $(d_i)$. Being that $d_i \leq n$, at most we will require to update $\tp_j$ and $\tx_j$, for $j \leq \min\{n,m\}$. This is specially relevant for infinite mixture models as it assures the sampler unfolds in a finite dimensional space even when the model dimension, $m$, is infinite. In fact, if the model dimension is deterministic, the ordered allocation sampler is practically identical for finite and infinite mixture models. The case of random $m$ is treated next.

\subsubsection*{Updating the model dimension $m$:}\label{sec:3.1}

If the model dimension is random, our proposal here is to update $m$, $(\tp_j)$ and the non occupied component parameters, $(\tx_j)_{j>k_n}=(\tx_{k_n+1},\ldots,\tx_m)$, as a block from 
\begin{multline}\label{eq:m,x,p_post_sb}
\pr[\,m,(\tp_j),(\tx_j)_{j>k_n}\mid\rest\,]
  \propto \prod_{j=1}^{k_n}\tp_j^{n_j-1}
  \bigg(1-\sum_{l=1}^{j-1}\tp_l\bigg)
  \prod_{j=k_n+1}^m \nu(\tx_j)\times\pr[\,(\tp_j)\mid m] \pr[m].
\end{multline}
Here we keep \comillas{$\rest$} to denote all random terms other than $m$, $(\tp_j)$ and $(\tx_j)_{j>k_n}$. We first sample $m$ from its {\it marginal}, i.e. \eqref{eq:m,x,p_post_sb} after integrating over $(\tp_j)$ and $(\tx_j)_{j>k_n}$:
\begin{equation}\label{eq:m_post_sb_0}
\pr[\,m\mid\rest\,]
\propto \Esp\bigg[\prod\nolimits_{j=1}^{k_n}\tp_j^{n_j-1} \bigg(1-\sum\nolimits_{l=1}^{j-1}\tp_l\bigg)\,\bigg|\, m\bigg] \pr(m).
\end{equation}
The expectation is taken with respect to the conditional distribution of $(\tp_j)$ given $m$ and treating $k_n,n_1,\ldots,n_{k_n}$ as constants. In particular, since $\sum_{j=1}^m \tp_j = 1$, \eqref{eq:m_post_sb_0} equals zero for $m < k_n$.
Taking this into account and recognizing, in the conditional expectation, the EPPF of the species sampling model given $m$, cf. \eqref{eq:eppf_sb}, we obtain
\begin{equation}\label{eq:m_post_sb}
\pr[\,m\mid \rest\,] \propto \pi(n_1,\ldots,n_{k_n}\mid m)\pr[m]\Ind_{\{k_n \leq m\}}.
\end{equation}
This is a remarkably simple expression for the updating of the model dimension as it only requires the conditional EPPF given $m$. In Appendix \ref{sec:supp_prior_m} we provide an example on how to update $m$ for mixtures of finite mixtures and for the choice of $\pr[m]$ detailed by \citet{Gne:10ecp}. 
After updating $m$, we sample $(\tp_j)$ and $(\tx_j)_{j>k_n}$, conditioning on $m$.
Thus $(\tp_j)$ is sampled from \eqref{eq:p_post_sb} as detailed before, and the $m-k_n$ non occupied component parameters, {$\tx_{k_n+1},\ldots,\tx_m$}, from the prior $\nu$, cf. \eqref{eq:x_post_sb}. Note that the blocking argument is remarkably simple when compared with the Metropolis-Hasting steps of the reversible jump MCMC algorithm. 



\subsection{Ordered allocation sampler for non size-biased weights}\label{sec:OAS}

In this section we adapt the ordered allocation sampler to species sampling priors that do not enjoy an explicit characterization of the size-biased weights $(\tp_j)$. This makes our sampler applicable to mixture models for which marginal samplers are not available, so increasing substantially its scope (examples can be found at the end of Section \ref{sec:2}). To this aim recall that $(\tp_j)$ is a rearrangement of the weights in any arbitrary order, $(p_j)$, i.e. $\tp_j = p_{\alpha_j}$, where $(\alpha_j)$ is sampled without replacement from $\{1,\ldots,m\}$ with probabilities $(p_j)$, as defined in \eqref{eq:size-biased_pick}. The key idea is to include $(\alpha_j)$ as part of the sampler.
As we will see, this augmentation yields a conditional sampler that inherits the advantages of the algorithm in Section \ref{sec:OASsb} without requiring a closed-form expression of $(\tp_j)$ or the EPPF.
As for the component parameters in order of appearance, $(\tx_j)$, we will continue to  model them directly, as i.i.d. from $\nu$, independently of $(p_j)$ and $(\alpha_j)$, cf. \emph{iv} in Theorem \ref{theo:rep_sss}. 
Thus, instead of \eqref{eq:like_sb}, we work with the augmented likelihood
\begin{equation}\label{eq:like_nsb}
\pr[\,(y_i),(d_i) \mid (p_j),(\alpha_j),(\tx_j)\,]
= \prod_{j=1}^{k_n}p_{\alpha_j}^{n_j-1}
  \bigg(1-\sum_{l=1}^{j-1}p_{\alpha_l}\bigg)
  \prod_{i \in D_j}g(y_i\mid\tx_{j})
  \Ind_{\mathcal{D}}.
\end{equation}
It is straightforward to see that the updating of the ordered allocation variables $(d_i)$ and the component parameters $(\tx_j)$ remain identical. Hence, we will only explain how to update the weights in order of appearance via $(p_j)$ and $(\alpha_j)$, as well as the model dimension, $m$, whenever this quantity is random.

\subsubsection*{Updating of $(\tp_j)$ through  $(p_j)$ and $(\alpha_j)$:}

From \eqref{eq:size-biased_pick} and \eqref{eq:like_nsb} we find
\begin{equation}\label{eq:p_a_post_nsb}
\pr[\,(p_j),(\alpha_j)\mid \rest\,] \propto \prod_{j=1}^{k_n}p_{\alpha_j}^{n_j} \times \prod_{j=k_n+1}^{m} p_{\alpha_j}\bigg(1-\sum_{l=1}^{j-1}  p_{\alpha_l}\bigg)^{-1}\Ind_{\clA} \times \pr[(p_j)],
\end{equation}
where $\clA$ is the event that $\alpha_i \neq \alpha_j$ for every $i \neq j$. 
In this part, as a notational device, we keep \comillas{$\rest$} to denote all random terms other than $(p_j),(\alpha_j)$. Also, we distinguish the indexes of the occupied components, $(\alpha_j)_{j\leq k_n}$, from the remaining ones, $(\alpha_j)_{j>k_n}$. 
The key idea to attain simple updating steps is to sample $(\alpha_j)_{j \leq k_n}$ from its full conditional, which can be expressed as a {\it weighted permutation} of $k_n$ indexes, and separately $(p_j)$ and $(\alpha_j)_{j>k_n}$ as a block. 

We first focus on the updating of $(\alpha_j)_{j \leq k_n}$. From  \eqref{eq:p_a_post_nsb} we get
\begin{equation}\label{eq:a_kn_post_nsb}
  \pr[\,(\alpha_j)_{j \leq k_n}\mid 
   (\alpha_j)_{j > k_n},(p_j),\rest\,] 
  \propto \prod_{j=1}^{k_n} p_{\alpha_j}^{n_j}\Ind_{\clA},
\end{equation}
after noting that $\prod_{j=k_n+1}^m(1-\sum_{l=1}^{j-1}p_{\alpha_l})^{-1}$ is a constant with respect to $(\alpha_j)_{j \leq k_n}$, because $1- \sum_{l=1}^{j-1}p_{\alpha_l} = \sum_{l=j}^{m}p_{\alpha_l}$. The event $\clA$ indicates that this is about sampling from a weighted permutation $\rho$ of the $k_n$ integers corresponding to current values of $(\alpha_{j})_{j\leq k_n}$.
Namely, we sample $\rho$ from
  $$\pi(\rho)=\frac1Z\prod_{j=1}^{k_n}w_{j,\rho(j)}\, , \quad
  \rho\in{\cal S},$$
where $w_{j,l}=p^{n_j}_{\alpha_l}$, $Z$ is the normalizing constant and ${\cal S}$ is the space of permutations of $\{1,\ldots,k_n\}$. Afterwards we simply apply $\rho$ to the indexes of the current value of $(\alpha_j)_{j \leq k_n}$ so to obtain the updated value $(\alpha_{\rho(j)})_{j \leq k_n}$.
Now, to sample $\rho$ from $\pi$ we follow \cite{Zan:20} by adopting a Metropolis–Hastings scheme using a {\it locally-balanced} informed proposal distribution, cf. Example 3 therein.
For the reader's convenience, we recall briefly how it works. Let $N(\rho)$ be the neighborhood of $\rho\in{\cal S}$ given by all permutations obtained by switching two indexes, (i.e. $\rho^* \in N(\rho)$ if and only if there exist $i\neq j$ such that $\rho^*(i) = \rho(j)$, $\rho^*(j) = \rho(i)$ and $\rho^*(l) = \rho(l)$ for all $l \not\in \{i,j\}$). 
Instead of using a random walk scheme, consisting in proposing a new value of $\rho$, say $\rho^*$, uniformly over $N(\rho)$, and accepting it with probability $a(\rho,\rho^*)=\min\{1,\pi(\rho^*)/\pi(\rho)\}$, we bias the proposal towards high probability regions of the target. To do so, we set the proposal distribution to be
  $Q(\rho,\rho^*)=\sqrt{\pi(\rho^*)}\Ind_{N(\rho)}/Z(\rho)$,
where 
  $Z(\rho)=\sum_{z\in N(\rho)}
  \sqrt{\pi(z)}$ 
is the normalizing constant. Then the new value, $\rho^*$, is accepted with probability 
  $a(\rho,\rho^*)=\min\big\{1,\frac{\pi(\rho^*)Q(\rho^*,\rho)}
  {\pi(\rho)Q(\rho,\rho^*)}\big\}$. 
In the simulation study we initialized $\rho$ as the identity function over ${\cal S}$ and performed $k_n$ Metropolis-Hastings steps at each iteration. As explained by \cite{Zan:20} the appeal of considering a locally-balanced proposal, such as $Q$, is that it is roughly $\pi$-reversible when ${\cal S}$ is large with respect to $N(\rho)$, thus the acceptance probability $a(\rho,\rho^*)$ tends to be high. Otherwise, if $k_n$ is small, one can opt to sample exactly from \eqref{eq:a_kn_post_nsb} by enumerating all possible permutations of $(\alpha_j)_{j \leq k_n}$.

As for the updating of $(p_j)$ and $(\alpha_j)_{j > k_n}$, we first sample $(p_j)$ from the conditional distribution $\pr[\,(p_j)\mid (\alpha_j)_{j \leq k_n},\rest \,]$ obtained from $\pr[\,(p_j),(\alpha_j)_{j > k_n} \mid (\alpha_j)_{j \leq k_n},\rest \,]$ after integrating over $(\alpha_j)_{j>k_n}$. We get
  $$
  \pr[\,(p_j)\mid (\alpha_j)_{j \leq k_n},\rest \,] 
  \propto \prod_{j=1}^{\overline{\alpha}}p_{j}^{r_j} 
  \times  \pr[(p_j)],
  $$
where $\overline{\alpha} = \max\{\alpha_j:j \leq k_n\}$, and $r_j = \sum_{l=1}^{k_n} n_l\Ind_{\{\alpha_l = j\}}$, that is $r_j = n_l$ if and only if there exist $l \leq k_n$ such that $\alpha_l = j$ and $r_j = 0$ otherwise. 
When $p_j = v_j \prod_{l=1}^{j-1} (1-v_l)$, we can update $(p_j)$ via sampling $(v_j)$ from
  $$
  \pr[\,(v_j)\mid (\alpha_j)_{j \leq k_n},\rest \,] 
 \propto \prod_{j=1}^{\overline{\alpha}}v_j^{r_j}(1-v_j)^{\sum_{l > j}r_j}\times \pr[(v_j)].
  $$ 
For instance, if a priori $v_j \sim \Be(a_j,b_j)$, independently for $j \geq 1$, then a posteriori $v_j \sim \Be(r_j+a_j,\sum_{l>j} r_l+b_j)$ for $j \leq \overline{\alpha}$ and $v_j \sim \Be(a_j,b_j)$, for $j >  \overline{\alpha}$. Further examples on how to update $(p_j)$ can be found in Appendix \ref{sec:supp_ESB}. After updating $(p_j)$ we sample $(\alpha_j)_{j > k_n}$ from
\begin{equation}\label{eq:a_star_post_nsb}
  \pr[\,(\alpha_j)_{j > k_n}\mid 
   (\alpha_j)_{j \leq k_n},(p_j),\rest\,] 
  = \prod_{j=k_n+1}^m p_{\alpha_j}\bigg(1-\sum_{l=1}^{j-1}  p_{\alpha_l}\bigg)^{-1}\Ind_{\clA}.
\end{equation}
That is, 
$\alpha_{k_n+1},\alpha_{k_n+2},\ldots$ are sampled without replacement from $\{j\leq m:\ j\not\in (\alpha_j)_{j \leq k_n}\}$ with probabilities proportional to $\{p_j:\ j\not\in (\alpha_j)_{j \leq k_n}\}$. This can be achieved by sampling  sequentially as a priori, cf. \eqref{eq:size-biased_pick}.

This way of updating $(\alpha_j)$, although theoretically valid, has the disadvantage that switches among indexes in $(\alpha_j)_{j \leq k_n}$ and indexes in $(\alpha_j)_{j > k_n}$ only occur when $k_n$ changes as a consequence of an update in $(d_i)$. To facilitate the mixing 
one can include the following {\it acceleration step} after updating $(\alpha_j)_{j \leq k_n}$ from \eqref{eq:a_kn_post_nsb} and before updating $(\alpha_j)_{j > k_n}$ from \eqref{eq:a_star_post_nsb}. We suggest to sample each $\alpha_j$ with $j \leq k_n$ from
\begin{equation}\label{eq:a_j_kn_post_nsb}
\pr[\,\alpha_j\mid (p_j),(\alpha_l)_{l \leq k_n, l \neq j}, \rest \,] \propto p_{\alpha_j}^{n_j}\Ind_{\clA},
\end{equation}
i.e. conditioning on the current values of $\alpha_l$, for $l \leq k_n$ and $l \neq j$, with $(\alpha_j)_{j>k_n}$ integrated out. 
Hence, the indicator $\Ind_{\clA}$ above only dictates $\alpha_j \neq \alpha_l$, with $l \leq k_n$. If the number of components is finite, the support of \eqref{eq:a_j_kn_post_nsb} consists of $m - k_n -1$ positive integers and sampling directly from this distribution is trivial. Otherwise, when $m = \infty$, we can treat $(\alpha_j)_{j > k_n}$ as a latent variable with distribution as in \eqref{eq:a_star_post_nsb}, and update the pair $(\alpha_j,\alpha_{k_n+1})$ from
\begin{equation}\label{eq:a_j_kn1_post_nsb}
\pr[\,(\alpha_j,\alpha_{k_n+1})\mid (p_j),(\alpha_l)_{l \not\in \{j,k_n+1\}}, \rest\,] \propto p^{n_j}_{\alpha_j}p_{\alpha_{k_n+1}} \bigg(1-\sum_{l=1}^{k_n}  p_{\alpha_l}\bigg)^{-1}\Ind_{\clA}.
\end{equation}
In practice, it is enough to sample $\alpha_{k_{n}+1}$ from $\pr[\,\alpha_{k_{n}+1}\mid \alpha_1,\ldots,\alpha_{k_n},(p_j),\rest\,]$ as in \eqref{eq:size-biased_pick} and later either leave $(\alpha_j,\alpha_{k_n+1})$ unchanged or switch the values of $\alpha_j$ and $\alpha_{k_n+1}$, with probabilities determined by \eqref{eq:a_j_kn1_post_nsb}. It is worth emphasizing that this procedure has to be repeated for all $j \leq k_n$, and that 
each time we discard $\alpha_{k_n+1}$ because it is only playing the role of an auxiliary variable to draw samples from \eqref{eq:a_j_kn_post_nsb}.

Similarly as with the sampler in Section \ref{sec:OASsb}, this sampler unfolds in a finite dimensional space even when the model dimension is infinite. In general, we will only need to update $\tx_j$ and $\alpha_j$, for $j > k_n$, when required by the updating the ordered allocation variables, $(d_i)$. At most iterations this will be necessary for only a few indexes $j$. As for the weights, it is the updating of $(\alpha_j)$ what will determine how many entries of $(p_j)$ must be updated. Thus, at most we will need to update $p_j$ for $j \leq \max\{\alpha_1,\ldots,\alpha_J\}$ where $J$ is the latest entry of $(\alpha_j)$ we were required to update.


\subsubsection*{Updating of $m$:}

If the model dimension is random, our proposal is to update $m$, $(\tx_j)_{j > k_n}$, $(p_j)$ and $(\alpha_j)_{j > k_n}$ as a block from the full conditional 
  $\pr[\,m,(\tp_j),(\tx_j)_{j>k_n}, (\alpha_j)_{j > k_n}
  \mid\rest\,]$.
Here we use \comillas{$\rest$} to denote all random terms other than $m$, $(\tx_j)_{j > k_n}$, $(p_j)$ and $(\alpha_j)_{j > k_n}$. Integrating over $(\tx_j)_{j > k_n}$, $(p_j)$  and $(\alpha_j)_{j > k_n}$, we first sample $m$ from
\begin{equation}\label{eq:m_post_nsb}
  \pr[\,m\mid\,\rest\,] \propto 
  \Esp\bigg[\prod\nolimits_{j=1}^{k_n}p_{\alpha_j}^{n_j}\,
  \bigg|\, m \bigg]\Ind_{\{m \geq k_n\}} \pr[m],
\end{equation}
where the expectation is taken with respect to the conditional distribution of $(p_j)$ given $m$, and treating $(n_j)_{j\leq k_n}$ and $(\alpha_j)_{j \leq k_n}$ as constants. 
Later we  sample $(\tp_j)$ and $(\alpha_j)_{j > k_n}$ from \eqref{eq:p_a_post_nsb} as previously explained, and the $m-k_n$ empty component parameters, $\tx_j$, from the prior, cf. \eqref{eq:x_post_sb}.
In contrast to \eqref{eq:m_post_sb}, the EPPF does not appear in \eqref{eq:m_post_nsb}, instead it is enough to compute an expectation of the weights. This is very convenient, being that when law of $(\tp_j)$ is not available, typically the EPPF is hard to compute as mentioned in Section \ref{sec:2}.

\subsection{Acceleration step}\label{sec:acc_steps}

There is a very simple modification of the ordered allocation sampler that can greatly improve its performance. 
To motivate it, first note that the set of admissible moves, $\D_i$, of $d_i$ is always contained in $\{1,\ldots,k_{i-1}+1\}$, with $k_0 = 0$ and $k_i = \max\{d_1,\ldots,d_{i}\}$. Recalling that $d_i$ indicates from which component of the mixture was $y_i$ sampled, this means that while the latest data points will be able to reallocate to virtually all observed components, the first data points will rarely be reassigned to a different component. Furthermore, since component parameters and weights are labelled in the order in which they were discovered by $(y_i)$, the initial order of data points can dictate how often there are label switches of components and thus affect the mixing properties of the sampler. To overcome this, it is enough to exploit the exchangeability of $(y_i)$ and add a step, after updating $(d_i)$, in which we randomly permute the data points obtaining 
$(y'_i) = (y_{\rho(i)})$, where $\rho$ is a uniform permutation of $\{1,\ldots,n\}$. Accordingly,
we modify $(d_i)$ obtaining $(d_i')$ defined by $d'_i = j$ if and only if $d_{\rho(i)}$ equals the $j$th distinct value to appear in $(d_{\rho(i)})$. This way,  the ordered allocation variables, $(d'_i)$, that correspond to the permuted data set, $(y'_i)$, preserve the induced clustering structure, and the least element order as dictated by the event $\D$ now holds for $(D'_j)$ with $D'_j = \{i:d'_i = j\}$ (see Appendix \ref{sec:supp_example_di} for an example). In accordance, for the sampler in Section \ref{sec:OAS} we will also need to change the values of $(\alpha_j)_{j\leq k_n}$ so to obtain $(\alpha'_j)_{j\leq k_n}$, where $\alpha'_j$ now indicates which weight in $(p_j)$ is the $j$th one to be discovered by $(y'_i)$. To do so we simply have to set $\alpha'_j = \alpha_l$ if and only if the $j$th distinct value to appear in $(d_{\rho(i)})$ equals $l$. After doing so we can move on with the updating of $m$ (if it is random) and each of the observed component parameters, $\tx'_j$, and weights, $\tp'_j = p_{\alpha'_j}$, identically as before, although now they are labelled in order in which they were discovered by $(y'_i)$. 

For the simulation study we will present in the following section, this acceleration step was included in all implementations of the ordered allocation. Nonetheless, in Appendix \ref{sec:supp_illust} we present a few runs of the ordered allocation sampler without it to illustrate its effect.


\section{Simulation study}\label{sec:4}

\begin{figure}
\centering
\includegraphics[width=1\textwidth]{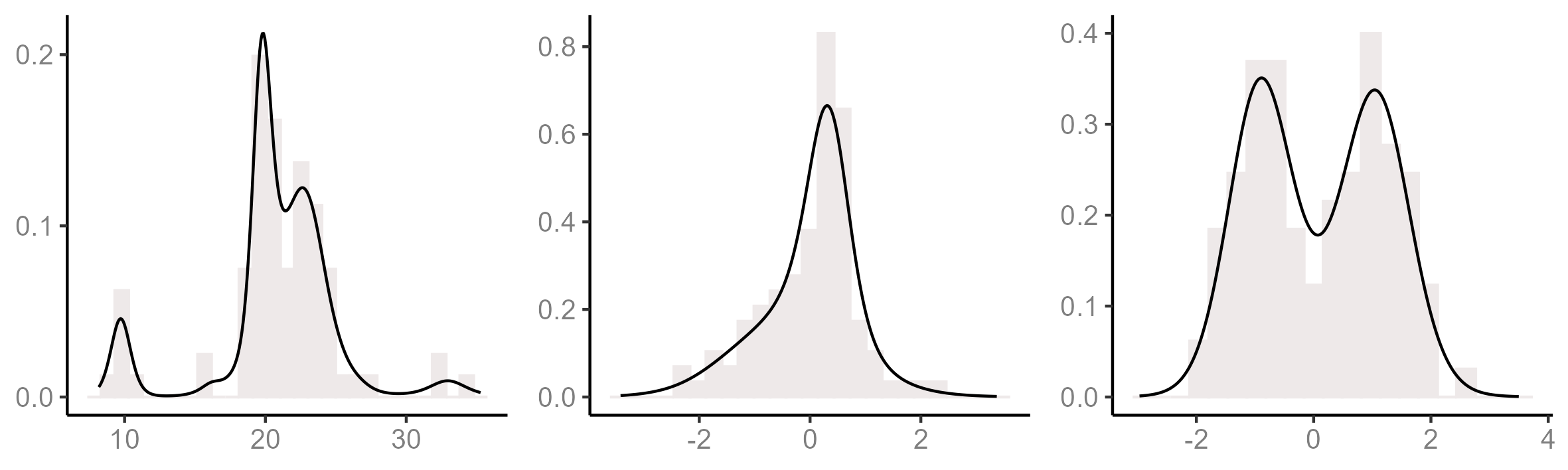}
\begin{small} 
\caption{Histogram of the $\mathsf{galaxy}$ (left), $\mathsf{leptokurtic}$ (middle) and $\mathsf{bimodal}$ (right) datasets. The lines represent the estimated densities using the ESB mixing prior and the ordered allocation sampler in Section \ref{sec:OAS}.
}\label{fig:data_ESB}
\end{small}
\end{figure}

In this section we present a simulation study to compare the mixing of the ordered allocation sampler against that of a marginal sampler and a conditional sampler. 
Following \cite{Kall:etal:11}, 
three different data sets have been considered (histograms are displayed in Figure \ref{fig:data_ESB}). The first data set is the $\mathsf{galaxy}$ data, consisting of the velocities of $82$ distinct galaxies diverging away from our galaxy. The other two data sets are the $\mathsf{leptokurtic}$ and $\mathsf{bimodal}$ data sets first introduced in \cite{Gre:Ric:01}. The $\mathsf{leptokurtic}$ consists of $100$ data points simulated from the mixture $0.67\,\mathsf{N}(0,1)+0.33\,\mathsf{N}(0.3,0.25^2)$. In the $\mathsf{bimodal}$ the $100$ observations come from the mixture $0.5\,\mathsf{N}(-1,0.5^2)+0.5\,\mathsf{N}(1,0.5^2)$. 
To each data set we fitted a mixture of Gaussian distributions with random location and scale parameters, i.e. $g(y\mid x) = \mathsf{N}(y\mid \mu,\sigma^2)$, and with five different mixing priors specifications. First we consider a mixture of finite mixtures \citep[MFM,][]{Mil:Har:18} specifically a mixing prior with random dimension, $m$, and symmetric Dirichlet weights, $(p_1,\ldots,p_m)\sim \Dir(\gamma,\ldots,\gamma)$, with $\gamma=1$. As for $m$, we used the prior $\pr[m] = \lambda(1-\lambda)_{m-1}/m!$ \citep{Gne:10ecp} with $\lambda = 0.1$. The remaining mixing priors we considered are a Dirichlet process (DP) with total mass parameter $\theta = 1$, a Pitman-Yor process (PY) with parameters $(\sigma,\theta) = (0.3,0.7)$, a Geometric process \citep[GP,][]{Fue:etal:10} and an Exchangeable Stick-Breaking process \citep[ESB,][]{Gil:Men:21}. Further specifications of the Geometric and the Exchangeable stick-breaking processes can be found in Appendix \ref{sec:supp_ESB}.
In all cases the distribution $\nu$ of the component parameters was fixed to $\nu(\mu,\sigma^2) = \mathsf{N}(\mu\mid \mu_0,\lambda_0^{-1}\sigma^2)\Gamma^{-1}(\sigma^2\mid a_0,b_0)$ with hyperparameters $\mu_0 = n^{-1}\sum_{i=1}^n y_i$, $\lambda_0 = 1/100$ and $a_0 = b_0 = 0.5$.


The marginal sampler we have implemented is Algorithm 8 in \cite{Nea:00} for DP, PY and MFM. In particular for MFM, Algorithm 8 was adapted following \cite{Mil:Har:18}. No marginal samplers are available for GP and ESB as these priors lack a P\'olya urn scheme representation. As for the conditional sampler, we implemented the (dependent) slice-efficient sampler, as described by \cite{Kall:etal:11}, for all models but MFM. The ordered allocation sampler (OAS in short) was used to implement all considered mixing priors. In particular,
we used the algorithm in Section \ref{sec:OASsb} for MFM, DP and PY, as these priors enjoy a tractable law of the size-biased permuted weights. The algorithm of Section \ref{sec:OAS} was used for GP and ESB. In Appendix \ref{sec:supp_ESB} we explain how to update the weights of the GP and ESB priors.

To monitor algorithmic performance we explored the convergence of  the number of occupied components, $k_n$, and the deviance, $D_v$, of the estimated density \citep[cf][]{Gre:Ric:01}. 
The deviance can be computed by
  $
  D_v = -2\sum_{i=1}^n\log
  \sum_{j=1}^{m}\frac{n_j}{n}g(y_i\mid x_j),
  $
where $n_j$ is the number of data points associated to $g(y_i\mid x_j)$. 
More precisely, 
we considered the chains 
$(k_n^{t})_{t=1}^T$ and $(D_v^{t})_{t =1}^{T}$ attained from $T$ iterations after the burn-in period.
In each case we estimated the integrated autocorrelation time  (IAT), $\tau = 1/2 + \sum_{l=1}^{\infty} \rho_l$, where $\rho_l$ stands for $l$-lag autocorrelation of the monitored chain. As done by \cite{Kall:etal:11}, $\tau$ was estimated through $\hat{\tau} = 1/2 + \sum_{l=1}^{C-1}\hat{\rho}_l$, where $\hat{\rho}_l$ is the estimated autocorrelation at lag $l$ and $C = \min\{l:|\hat{\rho}_l|< 2/\sqrt{T}\}$. This is a very useful summary statistic for quantifying the convergence of an MCMC algorithm, smaller values of $\hat{\tau}$ corresponding to better performance. 
For each sampler and mixing prior, we considered $2\times 10^6$ iterations after a burn-in period of $10^5$ iterations.
Table \ref{tab:all} reports
estimates of the IAT for $k_n$ and $D_v$ with standard errors appearing in parenthesis, the latter computed following Section 3 of \cite{Sok:97}.

\begin{table}
\begin{scriptsize}
\centering
\begin{tabular}{|c | c | C{1.2cm}C{1.2cm}C{1.55cm}|
                C{1.2cm}C{1.2cm}C{1.55cm}|}
\multicolumn{2}{c}{} & 
\multicolumn{3}{c}{$\mathsf{galaxy}$ data}&
\multicolumn{3}{c}{$\mathsf{leptokurtic}$ data}\\ 
\hline
\multicolumn{2}{|c}{} & Marginal & OAS  & Conditional 
& Marginal & OAS  & Conditional \\ \hline
\multirow{ 2}{*}{MFM} & $D_v$ 
& 14.43(0.38) & 26.17(0.93) & --- 
& 563.5(61.3) & 855.5(94.8) & --- \\
& $k_n$ 
& 33.55(0.92) & 89.42(3.07) & ---  
& 337.7(26.3) & 535.2(64.7) & --- \\ \hline
\multirow{ 2}{*}{DP} & $D_v$ 
& 12.30(0.23) & 23.76(0.57) & 119.2(9.95)
& 22.42(0.54) & 25.81(0.63) & 120.5(10.3) \\
& $k_n$ 
& 13.68(0.25) & 32.49(0.81) & 190.2(16.3)
& 9.26(0.13) & 18.99(0.41) & 100.8(7.18) \\ \hline
\multirow{ 2}{*}{PY} & $D_v$ 
& 13.48(0.24) & 21.59(0.52) & 83.33(4.63)
& 53.85(1.45) & 62.91(2.10) & 322.9(37.0) \\
& $k_n$ 
& 12.43(0.23) & 35.62(0.84) & 115.7(6.23)
& 12.02(0.24) & 20.96(0.56) & 138.7(13.1)\\ \hline
\multirow{ 2}{*}{GP} & $D_v$ 
& --- & 11.01(0.33) & 40.76(2.50)
& --- & 50.64(1.55) & 158.7(7.86) \\
& $k_n$ 
& --- & 61.67(1.89) & 621.2(172.5)
& --- & 45.34(1.21) & 129.8(6.43) \\ \hline
\multirow{ 2}{*}{ESB} & $D_v$ 
& --- & 24.29(0.68) & 245.4(28.3)
& --- & 61.29(1.97) & 184.5(18.6) \\
& $k_n$ 
& --- & 59.27(2.16) & 632.4(88.9)
& --- & 26.78(0.91) & 120.1(10.7) \\ \hline
\end{tabular}

\vspace{0.5cm}

\begin{tabular}{|c | c | C{1.2cm}C{1.2cm}C{1.55cm}|}
\multicolumn{2}{c}{} & 
\multicolumn{3}{c}{$\mathsf{bimodal}$ data} \\ 
\hline
\multicolumn{2}{|c}{}  
& Marginal & OAS  & Conditional \\ \hline
\multirow{ 2}{*}{MFM} & $D_v$ 
& 122.9(8.04) & 143.0(9.28) & --- \\
& $k_n$ 
& 65.15(3.89) & 109.4(7.99) & ---  \\ \hline
\multirow{ 2}{*}{DP} & $D_v$ 
& 7.84(0.16) & 13.87(0.35) & 35.61(1.68) \\
& $k_n$ 
& 6.30(0.07) & 13.38(0.22) & 52.00(2.18)  \\ \hline
\multirow{ 2}{*}{PY} & $D_v$ 
& 39.91(1.33) & 58.11(2.21) & 257.1(22.8) \\
& $k_n$ 
& 6.24(0.14) & 12.40(0.30) & 58.10(3.35)  \\ \hline
\multirow{ 2}{*}{GP} & $D_v$ 
& --- & 57.45(1.76) & 148.4(5.81) \\
& $k_n$ 
& --- & 55.85(1.65) & 146.0(5.94)  \\ \hline
\multirow{ 2}{*}{ESB} & $D_v$ 
& --- & 48.48(1.52) & 76.51(3.48) \\
& $k_n$ 
& --- & 19.93(0.58) & 35.53(1.42)  \\ \hline
\end{tabular}
\caption{Results for the three datasets by model and sampler.\label{tab:all}}
\end{scriptsize}
\end{table}

We observe that, when applicable, Algorithm 8 outperforms the other samplers, and that the OAS has better mixing properties than the slice sampler. On average, the IAT corresponding to the OAS is roughly $1.8$ times bigger than that of Algorithm 8, and the IAT of the slice sampler is approximately $4.7$ times larger than that of the OAS.
In general, it has been found that conditional algorithms perform worse than marginal samplers. This can be explained by the fact that in conditional algorithms mixing takes place in the space of all possible values of the (usual) allocation variables, $(c_i)_{i=1}^n$, given by $c_i = j$ if and only is $\theta_i = x_j$. Instead, in marginal algorithms the labels of the allocation variables are irrelevant, which means that the sampler searches is the space of partitions of $\{1,\ldots,n\}$, generated by the ties among allocation variables \citep[cf.][]{Port:etal:06}. Now, in the OAS, mixing occurs in the space of all possible values of the ordered allocation variables, $(d_i)$, which unequivocally define an ordered partition of $\{1,\ldots,n\}$, with blocks in the least element order. Since there exists a one to one correspondence between (unordered) partitions of $\{1,\ldots,n\}$ and partitions, of the same set, ordered according to the least element, we find that marginal samplers and the OAS search in the exact same space. This explains the better mixing properties of the OAS when compared with the slice sampler. Still, Algorithm 8 has better performances compared with the OAS, which is mainly due to the restricted support of $(d_i)$ in the OAS.

In Appendix \ref{sec:supp_illust} we extend this study for the DP model to further compare the distinct versions of the OAS in Sections \ref{sec:OASsb} and \ref{sec:OAS}, as well as the effect of the acceleration step in Section \ref{sec:acc_steps}. There we also show the graph of the estimated weighted densities by component, so to illustrate in more details how the different samplers mix over component labels.


\section{Discussion}\label{sec:5}


The ordered allocation sampler exploits the conditional law of a species sampling sequence given the atoms and the weights in order of appearance. The idea of sorting the parameters by order of appearance is analogous to that of \cite{Cho:07} for devising sequential Monte Carlo algorithm for hidden Markov models. A key difference is that in our framework we retain exchangeability of the data, while in hidden Markov model the data possess a precise temporal order.

Mixture models with a random dimension have been long known for their appeal from a modelling perspective and for their optimal asymptotic properties \citep{Rou:Men:11,She:etal:13}. However, posterior computation had remained somehow elusive until the advent of the marginal sampler by \citet{Mil:Har:18}. The ordered allocation sampler is a valid alternative, it is simple to implement,  and more broadly applicable. The sampler has been illustrated for mixtures of finite mixtures, but it readily applies to symmetric Dirichlet distributed weights whose parameter $\gamma$ can depend on the number of components. For example, we can use it to implement Dirichlet-multinomial mixing priors, or \comillas{sparse finite mixtures} as termed by \citet{Fru:Mal:19}, where $\gamma = \theta/m$. As for mixtures with infinitely many components, the sampler completely avoids the truncation problem. In fact, it is practically identical for the case where $m$ is a finite fixed number and the case where $m$ is infinite. 
Other conditional samplers are designed for the case where  $m < \infty$ is fixed, $m = \infty$, or $m$ is random and there are clear distinctions between samplers that are designed for one case or another. To the best of our knowledge, the ordered allocation sampler is the first conditional sampler that treats in a unified manner the distinct assumptions on $m$. 

As highlighted throughout the paper, the ordered allocation sampler enjoys nice properties in terms of applicability and mixing performance. Nonetheless, there are areas of improvements. In other conditional samplers allocation variables are updated independently of each other in a block, rather that one at a time. In big data settings this is a significant advantage over the ordered allocation sampler. Another drawback is that the number of occupied components, $k_n$, can not change as freely, from one iteration to the next one, as it does in other samplers. This  explains the higher IAT of $k_n$ when compared against the marginal method. 
The ordered allocation sampler may need additional modifications to address these issues, which is an interesting direction for future research.



\bibliographystyle{dcu}
\bibliography{short_biblio}

@String{JASA  = {{Journal of the American Statistical Association}}}

@String{JCGS  = {{Journal of Computational and Graphical Statistics}}}

@String{JRSSB = {{Journal of the Royal Statistical Society - Series B}}}

@String{JASA  = {{J. Amer. Statist. Assoc.}}}

@String{AOS   = {{Ann. Statist.}}}

@String{AOP   = {{Ann. Probab.}}}

@String{JCGS  = {{J. Comput. Graph. Statist.}}}

@String{JRSSB = {{J. R. Stat. Soc. Ser. B}}}

@INPROCEEDINGS{Sok:97,
  address = {Boston, MA},
  author = {Sokal, A.},
  booktitle = {Functional Integration: Basics and Applications},
  editor = {DeWitt-Morette, Cecile and Cartier, Pierre and Folacci, Antoine},
  pages = {131--192},
  publisher = {Springer US},
  title = {{Monte Carlo Methods in Statistical Mechanics: Foundations and New Algorithms}},
  year = {1997},
}

@article{Fav:Teh:13,
author = {Stefano Favaro and Yee Whye Teh},
title = {{MCMC} for normalized random measure mixture models},
volume = {28},
journal = {Statistical Science},
number = {3},
publisher = {Institute of Mathematical Statistics},
pages = {335 -- 359},
year = {2013}
}

@article{Bla:Mac:73,
author = {David Blackwell and James B. MacQueen},
title = {{Ferguson distributions via Polya urn schemes}},
volume = {1},
journal = AOS,
number = {2},
publisher = {Institute of Mathematical Statistics},
pages = {353 -- 355},
year = {1973},
doi = {10.1214/aos/1176342372}
}

@article{Gil:Men:21,
author = {María F. Gil–Leyva and Ramsés H. Mena},
title = {{S}tick-breaking processes with exchangeable length variables},
journal = {Journal of the American Statistical Association},
volume = {},
number = {},
pages = {in press},
year  = {2021},
publisher = {Taylor & Francis},
doi = {10.1080/01621459.2021.1941054},
}

@article{Fav:etal:16,
	author = {Favaro, S. and Lijoi, A. and Nava, C. and Nipoti, B. and Pr\"unster, I. and Teh, Y. W.},
	title= {{O}n the stick-breaking representation for homogeneous {NRMI}s}, 
    journal = {Bayesian Anal.},
	year = {2016},
    volume = {11},
    pages = {697--724},
    doi = {10.1214/15-BA964}
}

@article{Fru:Mal:19,
	author = {Fr\"uhwirth-Schnatter, S. and Malsiner-Walli},
	title= {From here to infinity: sparse finite versus Dirichlet process mixtures in model-based clustering}, 
    journal = {Advances in Data Analysis and Classification},
	year = {2019},
    volume = {13},
    pages = {33--64},
    doi = {10.1007/s11634-018-0329-y}
}

@article{Rod:Dun:11,
  author = {Rodr\'iguez, A. and Dunson, D. B.},
  title = {Nonparametric {B}ayesian models through probit stick-breaking processes},
  journal = {Bayesian Anal.},
  year = {2011},
  volume = {6},
  number = {1},
  pages = {145--178},
}

@article{Zan:20,
author = {Zanella, G.},
title = {Informed proposals for local {MCMC} in discrete spaces},
journal = JASA,
volume = {115},
number = {530},
pages = {852-865},
year  = {2020},
doi = {10.1080/01621459.2019.1585255}
}

@article{She:etal:13,
    author = {Weining Shen and Surya T. Tokdar and Subhashis Ghosal},
    title = {Adaptive {B}ayesian multivariate density estimation with {D}irichlet mixtures},
    journal = {Biometrika},
    year = {2013},
    volume = {100},
    issue = {3},
    pages = {623--640},
}

@article{Rou:Men:11,
author = {Rousseau, Judith and Mengersen, Kerrie},
title = {Asymptotic behaviour of the posterior distribution in overfitted mixture models},
journal = JRSSB,
volume = {73},
number = {5},
pages = {689-710},
year = {2011}
}

@article{Pit:Yor:97,
	author = {Pitman, Jim and Yor, Marc},
	title = {The two-parameter {P}oisson-{D}irichlet distribution derived from a stable subordinator},
	year = {1997},
	journal = AOP,
	volume = {25},
	Number = {2},
	pages = {855--900}
}

@article{Esc:Wes:95,
    author = {Escobar, Michael D. and West, Mike},
    title = {{B}ayesian density estimation and inference using mixtures},
    year = {1995},
    journal = JASA,
    volume = {90},
    pages = {577--588},
    CISid = {166426}
}

@article{Ric:Gre:97,
 author		= {Richardson, S and Green, P J},
 title		= {{On Bayesian analysis of mixtures with an unknown number of components}},
 journal	= JRSSB,
 volume		= {59},
 pages		= {731--792},
 number		= {},
 year		= {1997}
}

@article{Cho:07,
author = {Chopin, Nicolas},
title = {Inference and model choice for sequentially ordered hidden Markov models},
journal = {JRSSB},
volume = {69},
number = {2},
pages = {269-284},
year = {2007}
}

@article{Deb:etal:15,
	author = {De Blasi, Pierpaolo and Favaro, Stefano and Lijoi, Antonio and Mena, Ramses H and Pr{\"u}nster, Igor and Ruggiero, Matteo},
	title = {{Are Gibbs-type priors the most natural generalization of the Dirichlet process?}},
    year = {2015},
    journal = {IEEE Transactions on Pattern Analysis and Machine Intelligence},
    volume = {37},
    number = {},
    pages = {212--229}
}

@article{Fer:73,
  author = {Ferguson, T.S.},
  title = {A {B}ayesian analysis of some nonparametric problems},
  journal = AOS,
  year = {1973},
  volume = {1},
  number = {2},
  pages = {209--230},
}

@article{Fue:etal:10,
author = { R.   Fuentes-García  and  R. H.   Mena  and  S. G.   Walker },
title = {A new {B}ayesian nonparametric mixture model},
journal = {Communications in Statistics - Simulation and Computation},
volume = {39},
number = {4},
pages = {669-682},
year  = {2010},
publisher = {Taylor & Francis},
doi = {10.1080/03610910903580963},
eprint = { https://doi.org/10.1080/03610910903580963}
}

@article{Gne:10ecp,
author = {Alexander Gnedin},
title = {{A species sampling model with finitely many types}},
volume = {15},
journal = {Electronic Communications in Probability},
publisher = {Institute of Mathematical Statistics and Bernoulli Society},
pages = {79-88},
year = {2010},
doi = {10.1214/ECP.v15-1532}
}

@article{Gre:Ric:01,
 author		= {Green, P J and Richardson, S},
 title		= {{Modeling heterogeneity with and without the Dirichlet process}},
 journal	= {Scand. J. Stat.},
 volume		= {28},
 pages		= {355--375},
 number		= {},
 year		= {2001}
}

@article{Ish:Jam:01,
    author = {Ishwaran, H. and James, L. F.},
    title = {Gibbs sampling methods for stick-breaking priors},
    year = {2001},
    journal = {J. Amer. Statist. Assoc.},
    volume = {96},
    issue = {},
    pages = {161--173},
}

@book{Kall:05,
  author = {Kallenberg, O.},
  year = {2005},
  publisher = {Springer},
  edition = {First},
  title = {{Probabilistic Symmetries and Invariance Principles}},
  doi = {10.1007/0-387-28861-9}
}

@article{Kall:etal:11,
  author = {Kalli, M. and Griffin, J. E. and Walker, S.G.},
  title = {Slice sampling mixtures models},
  journal = {Statist. Comput.},
  year = {2011},
  volume = {21},
  pages = {93--105},
}

@article{Lij:Pru:Rig:20,
    author = {Lijoi, Antonio and Pr{\"u}nster, Igor and Rigon, Tommaso},
    title = "{The Pitman--Yor multinomial process for mixture modelling}",
    journal = {Biometrika},
    volume = {107},
    number = {4},
    pages = {891-906},
    year = {2020},
    month = {06},
    issn = {0006-3444},
    doi = {10.1093/biomet/asaa030}
}

@article{Mil:Har:18,
author = {Jeffrey W. Miller and Matthew T. Harrison},
title = {Mixture Models With a Prior on the Number of Components},
journal = JASA,
fjournal = "Journal of the American Statistical Association",
volume = {113},
number = {521},
pages = {340-356},
year  = {2018},
publisher = {Taylor & Francis},
doi = {10.1080/01621459.2016.1255636},
eprint = { https://doi.org/10.1080/01621459.2016.1255636}
}

@article{Nea:00,
    author = {Neal, Radford M.},
    title = {{Markov Chain Sampling Methods for {D}irichlet Process Mixture Models}},
    year = {2000},
    journal = JCGS,
    volume = {9},
    number = {2},
    pages = {249--265}
}

@article{Pap:Rob:08,
author = {Papaspiliopoulos, O. and Roberts, G. O.},
title = {{Retrospective Markov chain Monte Carlo methods for Dirichlet process hierarchical models}},
volume = {95},
pages = {169--186},
journal = {Biometrika},
year = {2008}
}

@INPROCEEDINGS{Pit:96ims,
  address = {Hayward, California},
  author = {Pitman, J.},
  booktitle = {Statistics, Probability and Game Theory; Papers in honor of {D}avid {B}lackwell},
  editor = {et al., T.S. Ferguson},
  pages = {245--267},
  publisher = {Institute of Mathematical Statistics},
  series = {Lecture Notes-Monograph Series},
  title = {{Some developments of the Blackwell-MacQueen urn scheme}},
  volume = {30},
  year = {1996},
}

@article{Pit:95,
  author = {Pitman, J.},
  title = {Exchangeable and partially exchangeable random partitions},
  journal = {Probab. {T}heory  {R}elat. {F}ields},
  volume = {102},
  number = {},
  pages = {145--158},
  year  = {1995},
}

@article{Pit:96,
  author = {Pitman, J.},
  title = {Random discrete distributions invariant under size-biased permutation},
  journal = {Adv. {A}ppl. {P}robab.},
  volume = {28},
  number = {2},
  pages = {525--539},
  year  = {1996},
}

@book{Pit:06,
  author = {Pitman, J.},
  title = {{Combinatorial Stochastic Processes}},
  volume = {1875},
  collection = {\'{E}cole d'\'et\'e de probabilit\'es de {S}aint-{F}lour, {XIII}---2002},
  series = {\'{E}cole d'\'et\'e de probabilit\'es de {S}aint-{F}lour},
  publisher = {Springer-{V}erlag {B}erlin {H}eidelberg},
  address = {New York},
  pages = {1--260},
  edition = {First},
  year = {2006},
}

@inproceedings{Port:etal:06,
    author = {Porteous, Ian and Ihler, Alex and Smyth, Padhraic and Welling, Max},
    title = {Gibbs sampling for (coupled) infinite mixture models in the stick breaking representation},
    year = {2006},
    booktitle = {Proceedings of the Twenty-Second Conference on Uncertainty in Artificial Intelligence (UAI2006)},
    pages = {385--392},
}

@article{Reg:etal:03,
  author = {Regazzini, E. and Lijoi, A. and Pr{\"{u}}nster, I},
  title = {Distributional results for means of normalized random measures with independent increments},
  journal = AOS,
  year = {2003},
  volume = {31},
  number = {2},
  pages = {560--585},
}

@article{Seth:94,
  author = {Sethuraman, J.},
  title = {{A constructive definition of Dirichlet priors}},
  journal = {Stat. Sin.},
  year = {1994},
  volume = {4},
  pages = {639--650},
}

@article{Walk:07,
  author = {Walker, S. G.},
  title = {Sampling the {D}irichlet mixture model with slices},
  journal = {Communications in {S}tatistics-{S}imulation and {C}omputation},
  year = {2007},
  volume = {36},
  number = {1},
  pages = {45--54},
}


\newpage
\setcounter{figure}{0}
\setcounter{table}{0}
\setcounter{equation}{0}
\renewcommand{\thefigure}{\thesection\arabic{figure}}
\renewcommand{\thetable}{\thesection\arabic{table}}
\renewcommand{\theequation}{\thesection\arabic{equation}}
\renewcommand{\thelemma}{\thesection.\arabic{lemma}}
\renewcommand{\thetheorem}{\thesection.\arabic{theorem}}
\renewcommand{\theremark}{\thesection.\arabic{remark}}

\appendix

\begin{center}
\begin{Large}
{\bf Appendix}
\end{Large}
\end{center}

\section{Proof of Theorem 1}\label{sec:supp_proof}

\begin{lemma}\label{lem:peppf}
Let $m$ be a random variable taking values in $\{1,2,\ldots\}\cup\{\infty\}$ and let $(\tp_j) = (\tp_1,\ldots,\tp_m)$ be a sequence in $[0,1]$ with $\sum_{j=1}^m \tp_j = 1$. Let $\pi:\bigcup_{k \in \N}\N^{k} \to [0,1]$ be defined by
\[
\pi(n_1,\ldots,n_k) = \Esp\bigg[\prod_{j=1}^k \tp_j^{n_j-1}\prod_{j=1}^{k-1}\bigg(1-\sum_{l=1}^j \tp_l\bigg)\bigg].
\]
Then $(\tp_j)$ is invariant under size-biased permutations if and only if $\pi$ is a symmetric function of $(n_1,\ldots,n_k)$. 
\end{lemma}
The proof of Lemma \ref{lem:peppf} can be found in \cite{Pit:95,Pit:96}. Actually, Pitman derived it more in general for a sequence $(\tp_j)_{j=1}^{\infty}$ taking values in the infinite dimensional simplex $\Delta_{\infty} = \l\{(w_j)_{j=1}^{\infty}: w_j \geq 0, \sum_{j=1}^{\infty}w_j = 1\r\}$. The statement in Lemma \ref{lem:peppf} easily follows by transforming $(\tp_1,\ldots,\tp_m)$ into a sequence in $\Delta_{\infty}$ by appending zeros, i.e. $\tp_j = 0$ for $j > m$. For simplicity we will first take for granted Lemma \ref{lem:peppf}, later in Remark \ref{rem:self_cont} we explain how to derive a self-contained proof.

\subsection*{Proof of Theorem 1:}

(Sufficiency): Assume $(\theta_i)$ is a species sampling sequence driven by the species sampling model $P = \sum_{j=1}^m p_j\delta_{x_j}$. By de Finetti's theorem,
\[
\lim_{n \to \infty} \frac{1}{n}\sum_{i=1}^n \delta_{\theta_i} = \sum_{j=1}^m p_j\delta_{x_j},
\]
almost surely. As $\sum_{j=1}^m p_j = 1$ and $p_j > 0$, we get that outside a $\Prob$-null event, $\theta_i \in \{x_1,\ldots,x_m\}$ for every $i \geq 1$, and for each $j \leq m$ there exist $i \geq 1$ such that $\theta_i = x_j$. This together with the diffuseness of $\nu$, yield that $(\theta_i)$ exhibits exactly $m$ distinct values, $x_1,\ldots,x_m$, almost surely. This means that we can define $(\alpha_j) = (\alpha_1,\ldots,\alpha_m)$ given by  $\alpha_l = j$ if and only if $\theta_{M_l} = x_j$, recalling that  $M_1 =1$ and for $2 \leq l \leq m$, $M_l =  \min\{i > M_{l-1} :  \theta_i \not\in \{\theta_{M_1},\ldots,\theta_{M_{l-1}}\}\}$. This way, $\tilde{x}_j = x_{\alpha_j}$ is the $j$th distinct value of $(\theta_i)$ in order of appearance. Next we prove that $(\alpha_j)$ satisfies equation \eqref{eq:size-biased_pick} in the main document. To this aim, note that $P$ is $\{(p_j),(x_j)\}$-measurable and vice versa.  Moreover $(\theta_{M_1},\ldots,\theta_{M_l})$ is $\{(\alpha_1,\dots,\alpha_{l}),(x_j)\}$-measurable and $(\alpha_1,\dots,\alpha_{l})$ is $\{(\theta_{M_1},\ldots,\theta_{M_l}),(x_j)\}$-measurable. As $(\theta_i)$ is conditionally iid from $P$, this implies
\[
\Prob[\alpha_1 = j\mid (p_j),(x_j)] = \Prob[\theta_1 = x_j \mid P] = p_j
\]
and for $2 \leq l \leq m$,
\begin{align*}
\Prob[\alpha_l = j\mid (p_j),(x_j),\alpha_1,\ldots,\alpha_{l-1}]& = \Prob[\theta_{M_l} = x_j \mid P,\theta_{M_1},\ldots,\theta_{M_{l-1}}]\\
& \propto p_j\Ind_{\{x_j\not\in \{\theta_{M_1},\ldots,\theta_{M_{l-1}}\}\}}
\end{align*}
This is
\[
\Prob[\alpha_l = j\mid m,(p_j),(x_j),\alpha_1,\ldots,\alpha_{l-1}] = \frac{p_j}{1-\sum_{i=1}^{l-1}p_{\alpha_i}}\Ind_{\{j\not\in \{\alpha_1,\ldots,\alpha_{l-1}\}\}}.
\]
Hence, given $(p_j)$, $(\alpha_j)$ satisfies \eqref{eq:size-biased_pick}. Moreover, as $(p_j)$ is independent of $(x_j)$, we also get that $(\alpha_j)$ is independent of $(x_j)$, which are iid from $\nu$. This yields that $(\tx_j) = (x_{\alpha_j})$ are also iid from $\nu$, so \emph{i} is proved.

Using de Finetti's theorem once more, we get
\[
\lim_{n \to \infty}\sum_{j=1}^{k_n}\frac{|\{i \leq n: \theta_i = \tilde{x}_j\}|}{n}\delta_{\tilde{x}_j} = \lim_{n \to \infty}\frac{1}{n}\sum_{i=1}^{n}\delta_{\theta_i} = \sum_{j = 1}^m p_j \delta_{x_j} = \sum_{j = 1}^m p_{\alpha_j} \delta_{x_{\alpha_j}},
\]
where $k_n$ is the number of distinct values in $\{\theta_1,\ldots,\theta_n\}$. Since the directing random measure of an exchangeable sequence is unique almost surely, this assures $k_n \to m$, and for $j \leq m$, the long run proportion of indexes $i$ such that $\theta_i = \tilde{x}_j = x_{\alpha_j}$ is
\[
\tp_j = \lim_{n \to \infty}\frac{|\{i \leq n: \theta_i = \tilde{x}_j\}|}{n} = p_{\alpha_j},
\]
almost surely. As \eqref{eq:size-biased_pick} holds for $(\alpha_j)$, $(\tilde{p}_j) =({p}_{\alpha_j})$ is a size-biased permutation of $(p_j)$, which yields \emph{ii}. 

As for \emph{iii}, first note that by definition, $\tx_1,\ldots,\tx_{k_n}$ are the $k_n$ distinct values in order of appearance in $\{\theta_1,\ldots,\theta_n\}$, for every $n \geq 1$, in particular $\theta_1 = \tx_1$. Now, as $(\theta_i)$ is conditionally iid from $P = \sum_{j=1}^m p_j \delta_{x_j} = \sum_{j=1}^m \tp_j \delta_{\tx_j}$,  we get that for each $n \geq 1$ and every $j \leq k_n$,
\[
\Prob[\theta_{n+1} = \tx_j \mid (\tp_j),(\tx_j),\theta_1,\ldots,\theta_n] = \tp_j.
\]
Thus,
\[
\Prob[\theta_{n+1} \not\in \{\tx_1,\ldots,\tx_{k_n}\} \mid  (\tp_j),(\tx_j),\theta_1,\ldots,\theta_n] = 1-\sum_{j=1}^{k_n}\tp_j.
\]
By definition, under the event $\theta_{n+1} \not\in \{\tx_1,\ldots,\tx_{k_n}\}$ we must have $\theta_{n+1} = \tx_{k_n +1}$, i.e.
\[
\Prob[\theta_{n+1} \in \cdot \mid (\tx_j),(\tp_j),\theta_1,\ldots,\theta_n] = \sum_{j=1}^{k_n}\tilde{p}_j\delta_{\tilde{x}_j} + \bigg(1- \sum_{j=1}^{k_n}\tilde{p}_j\bigg)\delta_{\tilde{x}_{k_n +1}}.
\]

As for \emph{iv}, first note that $(\tp_j)$ is $\{(p_j),(\alpha_j)\}$-measurable and $(p_j)$ is $\{(\tp_j),(\alpha_j)\}$-measurable. The last assertion relies on $p_j = \tp_{\alpha^{-1}_j}$, where $(\alpha^{-1}_j)$ is the inverse permutation of $(\alpha_j)$. From the proof of \emph{i} and the hypothesis, we have that $(\alpha_j)$, $(p_j)$ and $m$ are independent of $(x_j)$. Thus, for every $B \in \B(\X)$,
\[
\Prob[\tx_j \in B \mid m,(\tp_j),(\alpha_j)] = \Prob[\tx_j \in B \mid m,(p_j),(\alpha_j)] = \Prob[x_{\alpha_j} \in B \mid m,(p_j),(\alpha_j)] = \nu(B),
\]
which proves \emph{iv}.

(Necessity): Assume \emph{i}--\emph{iv} hold. We first prove that $(\theta_i)$ is exchangeable. Fix $n \geq 1$ and define the random partition, $\Pi_n$ of $[n] = \{1,\ldots,n\}$ generated by the random equivalence relation $i \sim j$ if and only of $\theta_i = \theta_j$. In other words $\Pi_n = \{D_1,\ldots,D_{k_n}\}$ where $D_j = \{i \leq n: \theta_i = \tx_j\}$. Using \emph{iii}, a simple counting argument implies
\begin{equation}\label{eq:Pi_n_giv_m,p,x}
\Prob[\Pi_n = \{A_1,\ldots,A_k\}\mid (\tp_j),(\tx_j)] = \prod_{j=1}^k \tp_j^{n_j-1}\prod_{j=1}^{k-1}\bigg(1-\sum_{l=1}^j \tp_j\bigg),
\end{equation}
for every partition $\{A_1,\ldots,A_k\}$ of $[n]$, and where $n_j = |A_j|$. Taking expectations in \eqref{eq:Pi_n_giv_m,p,x}, 
\[
\Prob[\Pi_n = \{A_1,\ldots,A_k\}] = \Esp\bigg[\prod_{j=1}^k \tp_j^{n_j-1}\prod_{j=1}^{k-1}\bigg(1-\sum_{l=1}^j \tp_j\bigg)\bigg].
\]
By \emph{ii} and Lemma \ref{lem:peppf}, the function
\[
(n_1,\ldots,n_k) \mapsto \pi(n_1,\ldots,n_k) = \Esp\bigg[\prod_{j=1}^k \tp_j^{n_j-1}\prod_{j=1}^{k-1}\bigg(1-\sum_{l=1}^j \tp_j\bigg)\bigg]
\]
is symmetric. This shows $\Prob[\Pi_n = \{A_1,\ldots,A_k\}] = \pi(n_1,\ldots,n_k)$, at most depends on the number of blocks $k$ of $\{A_1,\ldots,A_k\}$ and the frequencies, $n_1,\ldots,n_k$, of each block, through a symmetric function. In other words, $\Pi_n$ is exchangeable, in the sense that for every permutation, $\rho$, of $[n]$, $\Pi_n$ is equal in distribution to $\rho(\Pi_n)$, where
\[
\rho(\Pi_n) = \{\rho(D_j): D_j \in \Pi_n\} \quad \text{ and } \quad \rho(D_j) = \{\rho(i): i \in D_j\}.
\]
Now, fix $B_1,\ldots,B_n \in \B(\X)$ and note
\begin{align*}
\Prob[\theta_{\rho(1)} \in B_1,\ldots,\theta_{\rho(n)} \in B_n\mid \Pi_n] & = \Prob[\tx_j \in B_l, \forall \, l \in \rho(D_j), j \leq k_n\mid \Pi_n]
 = \prod_{j=1}^{k_n} \nu\bigg(\bigcap_{i \in \rho(D_j)}B_i\bigg).
\end{align*}
The last equality follows from the fact that $(\tx_j)$ are iid from $\nu$, and $\tx_j$ is independent of $m$ and $(\tp_j)$, which together with \eqref{eq:Pi_n_giv_m,p,x} imply $\tx_j$ is independent of $\Pi_n$. By taking expectations in the last equation we find,
\begin{equation*}
\Prob[\theta_{\rho(1)} \in B_1,\ldots,\theta_{\rho(n)} \in B_n] = \Esp\bigg[\prod_{j=1}^{k_n} \nu\bigg(\bigcap_{i \in \rho(D_j)}B_i\bigg)\bigg].
\end{equation*}
As $\Pi_n$ is exchangeable,
\[
\Esp\bigg[\prod_{j=1}^{k_n} \nu\bigg(\bigcap_{i \in \rho(D_j)}B_i\bigg)\bigg] = \Esp\bigg[\prod_{j=1}^{k_n} \nu\bigg(\bigcap_{i \in D_j}B_i\bigg)\bigg],
\]
hence
\[
\Prob[\theta_{\rho(1)} \in B_1,\ldots,\theta_{\rho(n)} \in B_n]
= \Prob[\theta_1 \in B_1,\ldots,\theta_n \in B_n],
\]
which proves $(\theta_i)$ is exchangeable. Finally, by de Finetti's theorem we know that directing random measure of $(\theta_i)$ is given by
\[
P = \lim_{n \to \infty}\frac{1}{n}\sum_{i=1}^{n}\delta_{\theta_i} = \lim_{n \to \infty}\frac{|\{i \leq n: \theta_i = \tilde{x}_j\}|}{n}\sum_{j=1}^{k_n}\delta_{\tilde{x}_j},
\]
and by \emph{i} and \emph{ii} we conclude
\[
P = \sum_{j=1}^m \tp_j \delta_{\tx_j} = \sum_{j=1}^m p_{\alpha_j} \delta_{x_{\alpha_j}} = \sum_{j=1}^m p_j \delta_{x_j}.
\]
\qed

\begin{remark}\label{rem:self_cont}
The sufficiency of Lemma \ref{lem:peppf}, which we require to prove the necessity of Theorem 1, can be easily derived using the sufficiency of Theorem 1, thus provide a self-contained proof of Theorem 1. Namely, in the context of Lemma \ref{lem:peppf} let $(\tp_j)$ be invariant under size-biased permutations, and let $(p'_j) = (p'_1,\ldots,p'_{m'})$ be any sequence of weights whose size-biased permutation has the law of $(\tp_j)$. Then we can construct a species sampling model $P' = \sum_{j=1}^{m'} p'_j \delta_{x'_j}$ over a Borel space $(\X,\B(\X))$ and a species sampling sequence $(\theta'_i)$ driven by $P'$. By the sufficiency of Theorem 1, we get that for every $n \geq 1$,
\begin{equation}\label{eq:theta_peppf}
\Prob[\theta'_{n+1}\in \cdot\mid  (\tp'_j),(\tx'_j),\theta'_1,\ldots,\theta'_n]= \sum_{j=1}^{k'_n}\tp'_j\delta_{\tx'_j} + \bigg(1- \sum_{j=1}^{k_n}\tp'_j\bigg)\delta_{\tx'_{k'_n +1}},
\end{equation}
where $(\tx'_j)$ are the distinct values that $(\theta'_i)$ exhibits in order of appearance, and $(\tp'_j)$ denotes the size-biased permutation of $(p'_j)$. Now, let $\Pi'_n$ be the random partition of $[n]$ generated by the random equivalence relation $i \sim j$ if and only if $\theta'_i = \theta'_j$. Then, by construction $\Pi'_n$ is exchangeable, and a simple counting argument, using \eqref{eq:theta_peppf}, yields 
\[
\Prob[\Pi'_n = \{A_1,\ldots,A_k\}] = \Esp\bigg[\prod_{j=1}^k \bigg(\tp'_j\bigg)^{n_j-1}\prod_{j=1}^{k-1}\bigg(1-\sum_{l=1}^j \tp'_j\bigg)\bigg],
\]
where $n_j = |A_j|$. Since $(\tp'_j)$ is equal in distribution to $(\tp_j)$ and $\Pi'_n$ is exchangeable, we conclude
\[
\pi(n_1,\ldots,n_k) = \Esp\bigg[\prod_{j=1}^k \tp_j^{n_j-1}\prod_{j=1}^{k-1}\bigg(1-\sum_{l=1}^j \tp_j\bigg)\bigg] = \Esp\bigg[\prod_{j=1}^k \bigg(\tp'_j\bigg)^{n_j-1}\prod_{j=1}^{k-1}\bigg(1-\sum_{l=1}^j \tp'_j\bigg)\bigg]
\]
is a symmetric function of $(n_1,\ldots,n_k)$.
\end{remark}




\section{Mixtures of finite mixtures with \citet{Gne:10ecp} prior on $m$}\label{sec:supp_prior_m}

In this section we illustrate how to update the model dimension $m$ by sampling from \eqref{eq:m_post_sb}. We consider a mixture model with symmetric Dirichlet weights $(p_1,\ldots,p_m) \sim \Dir(1,\ldots,1)$, and random $m$ with prior distribution
\[
\pr[m] = \frac{\lambda(1-\lambda)_{m-1}}{m!},
\]
where $\lambda \in (0,1)$ is a known constant. As mentioned in Section \ref{sec:2}, the size-biased permuted weights $(\tp_j)$ admit stick-breaking representation $\tp_1 = v_1$, and $\tp_j = v_j\prod_{i=1}^{j-1}(1-v_i)$  for independent random variables, $v_j \sim \Be(1-\sigma,\theta+j\sigma)$ where $\sigma = -1$ and $\theta = m$. Thus, the ordered allocation sampler as derived in Section \ref{sec:OASsb} can be used to implement this model. 

First note that using \eqref{eq:eppf_sb} and the stick-breaking decomposition of $(\tp_j)$, we can compute the conditional EPPF given $m$:
\[
\pi(n_1,\ldots,n_{k_n}\mid m) = \frac{\prod_{j=1}^{k_n-1}(\theta+j\sigma)}{(\theta+1)_{n-1}}\prod_{j=1}^{k_n}(1-\sigma)_{n_j-1} = \frac{(m-k_n+1)_{k_n-1}}{(m+1)_{n-1}}\prod_{j=1}^{k_n}n_j!
\]
where $n = \sum_{j=1}^{k_n} n_j$ and $(z)_r = \prod_{i=0}^{r-1}(z+i)$, using the convention that the empty product equals one. Hence, \eqref{eq:m_post_sb} simplifies to
\[
\pr[\,m\mid \rest\,] \propto \frac{(m-k_n+1)_{k_n-1}}{(m+1)_{n-1}} \times \frac{\lambda(1-\lambda)_{m-1}}{m!} \Ind_{\{k_n \leq m\}}.
\]
Following \cite{Gne:10ecp}, we obtain
\[
\sum_{m = k_n}^{\infty}  \frac{(m-k_n+1)_{k_n-1}}{(m+1)_{n-1}} \frac{\lambda(1-\lambda)_{m-1}}{m!} = \frac{(k_n-1)!(1-\lambda)_{k_n-1}(\lambda)_{n-k_n}}{(n-1)!(1+\lambda)_{n-1}}.
\]
Thus, we can explicitly compute
\begin{align*}
\pr[\,m\mid &\rest\,] = \frac{{\lambda}(1-\lambda)_{m-1}(m-k_n+1)_{k_n-1}}{(m+n-1)!} \times\frac{(n-1)!(1+\lambda)_{n-1}}{(k_n-1)!(1-\lambda)_{k_n-1}(\lambda)_{n-k_n}}\Ind_{\{k_n \leq m\}}.
\end{align*}
In particular, using notation $q_{r} = \pr[\,m = r\mid \rest \,]$
\[
q_{k_n}= \frac{(\lambda+n-k_n)_{k_n}}{(n)_{k_n}},
\]
and recursively for $r \geq k_n$,
\[
q_{r+1} = q_{r}\frac{r(r-\lambda)}{(r-k_n+1)(r+n)}.
\]
Thus, to update $m$, sample $u \sim \mathsf{Unif}(0,1)$, and set $m = r$ when 
  $\sum_{l=k_n}^{r-1} q_l < u \leq \sum_{l=k_n}^{r} q_l$.


\section{Geometric and exchangeable stick-breaking processes}\label{sec:supp_ESB}

To illustrate the ordered allocation sampler derived in Section \ref{sec:OAS} we chose two species sampling mixing priors for which the law of $(\tp_j)$ is not available.
These are the geometric process and the exchangeable stick-breaking process. The geometric process \citep{Fue:etal:10} is a species sampling model $P = \sum_{j=1}^{\infty} p_j \delta_{x_j}$ with decreasingly ordered weights, $(p_j)$, given by $p_j = v(1-v)^{j-1}$ where $v$ is a random variable taking values in $(0,1)$.
The exchangeable stick-breaking process \citep{Gil:Men:21}
instead has weights $p_j = v_j\prod_{l=1}^{j-1}(1-v_l)$, where $(v_j) = (v_j)_{j=1}^{\infty}$ is an exchangeable sequence with values in $(0,1)$. Here we consider $(v_j)$ to be a species sampling sequence driven by a Dirichlet process $P'$ over $([0,1],\B([0,1]))$, with total mass parameter $\theta'$ and base measure $\nu' = \Be(a,b)$. 
Next we refer to $P = \sum_{j=1}^{\infty} p_j \delta_{x_j}$ as Dirichlet driven exchangeable stick-breaking process.

To fully specialize the ordered allocation sampler in Section \ref{sec:OAS} for this two mixing priors
it is enough to explain how to update $(p_j)$ via sampling $(v_j)$ from
\begin{equation}\label{eq:post_v}
\pr[\,(v_j) \mid \rest\,] \propto \prod_{j=1}^{\oal}v_j^{r_j}(1-v_j)^{R_j}\pr[(v_j)],
\end{equation}
where $r_j = \sum_{l=1}^{k_n} n_l\Ind_{\{\alpha_l = j\}}$, 
$R_j = \sum_{l > j}r_l$, $\overline{\alpha} = \max\{\alpha_1,\ldots,\alpha_{k_n}\}$, and \comillas{$\rest$} refers to all the random variables involved excluding $(\alpha_j)_{j > k_n}$ and $(p_j)$. Note that by excluding $(p_j)$ we are also excluding $(v_j)$ because these two sequences characterize each other. It is worth noting that the following description can be readily adapted to the updating of $(v_j)$ in the slice-efficient sampler.

For the geometric process we have that $v_j = v$ for every $j \geq 1$, hence it suffices to update $v$ from
\[
\pr[\,v\mid\rest\,] \propto v^n (1-v)^{\sum_{i=1}^n c_i - n} \times \pr[v],
\]
where $c_i = \alpha_{d_i}$ for each $i \leq n$. In particular if $v \sim \Be(a,b)$ a priori, then we update $v \sim \Be(a+n,b+\sum_{i=1}^nc_i -n)$. 

Now, for Dirichlet driven exchangeable stick-breaking processes, the updating of $(v_j)$ is more delicate due to the non-trivial dependence among elements in $(v_j)$. We will first focus on updating $(v_j)_{j \leq \overline{\alpha}}$. To this aim note that since $(v_j)$ is a species sampling sequence driven by a Dirichlet process with total mass parameter $\theta'$ and base measure $\nu' = \Be(a,b)$, we can compute
\[
\pr[(v_j)_{j \leq \overline{\alpha}}] = \frac{(\theta')^{k_{\oal}}}{(\theta')_{n}}\prod_{l=1}^{k_{\oal}}(m_l-1)!\Be(v^*_l\mid a,b),
\]
where $(v^*_l) = (v^*_1,\ldots,v^*_{k_{\oal}})$ are the distinct values that $(v_j)_{j \leq \oal}$ exhibits, $m_l = |E_l|$ and $E_l = \{j \leq \oal: v_j = v^*_l\}  =\{j \leq \oal: e_j = l\}$, with $e_j = l$ if and only if $v_j = v^*_l$ \citep[cf.][]{Pit:96ims,Nea:00}. Thus \eqref{eq:post_v} yields
\[
\pr[\,(v^*_l),(e_j) \mid \rest\,] \propto  \frac{(\theta')^{k_{\oal}}}{(\theta')_{n}}\prod_{l=1}^{k_{\oal}}(m_l-1)!(v^*_l)^{\sum_{j \in E_l}r_j}(1-v^*_l)^{\sum_{j \in E_l}R_j}\Be(v^*_l\mid a,b).
\]
Now to update $(v_j)$ we can first sample $(v^*_l)$ from
\[
\pr[\,(v^*_l)\mid (e_j) , \rest\,] \propto \prod_{l=1}^{k_{\oal}}\Be\bigg(v^*_l \,\bigg|\, a + \sum_{j \in E_l}r_j,b+\sum_{j \in E_l}R_j\bigg),
\]
which is a product of independent Beta distributions. Afterwards, for each $j \leq \oal$, we can update which value does $v_j$ take among the ones observed in the rest of the $v_j$'s or if it takes a new unobserved value. Say that $(v^*_l)_{-j} = (v^*_1,\ldots,v^*_{k_{-j}})$ are the distinct values in $(v_i:i \leq \oal, i \neq j)$, in no particular order, and assume without loss of generality that $e_i = l$ if and only if $v_i = v^*_l$ for each $i \neq l$. Then it is enough to sample from
\[
\pr[e_j = e \mid (e_i)_{i \neq j}, (v^*_l)_{-j}, \rest\,] \propto \begin{cases}
m_e (v^*_e)^{r_j}(1-v^*_e)^{R_j} &\text{ if } e \in \{1,\ldots,k_{-j}\}\\
\theta'\int v^{r_j}(1-v)^{R_j} \Be(dv\mid a,b)&\text{ if } e = k_{-j}+1\\
0 &\text{otherwise},
\end{cases}
\]
where $m_e = |\{i \neq j: e_i = e\}|$ and 
\[
\int v^{r_j}(1-v)^{R_j} \Be(dv\mid a,b) = \frac{\Gamma(r_j+a)\Gamma(R_j+b)}{\Gamma(r_j+R_j+a+b)}. 
\]
If the updated value $e_j \in \{1,\ldots,k_{-j}\}$ we simply set $v_j = v^*_{e_j}$ otherwise if $e_j = k_{-j}+1$ we sample $v_j \sim \Be(a+r_j,b+R_j)$. Once we have updated $(v_j)_{j \leq \oal}$, we can update $v_j$ for $j > \oal$ by sampling sequentially from $\pr[\,v_j\mid (v_i)_{i < j},\rest\,]$, which happens to coincide with the prior prediction rule $\pr[\,v_j\mid (v_i)_{i < j}\,]$. As $(v_j)$ is a species sampling sequence driven by a Dirichlet process, $P'$, with total mass parameter $\theta'$ and base measure $\nu' = \Be(a,b)$, it is well known that
\[
\Prob[v_j \in \cdot \mid (v_i)_{i < j}] = \sum_{j=1}^{k_{j-1}}\frac{m_l}{\theta'+j-1}\delta_{v^*_l} + \frac{\theta'}{\theta'+j-1}\nu'
\]
where $v^*_1,\ldots,v^*_{k_{j-1}}$ are the distinct values in $(v_i)_{i < j}$, and $m_l = |\{i < j: v_i = v^*_l\}|$ \cite[cf.][]{Pit:96ims}. Thus updating $v_j$ for $j > \oal$ is easy, and it will be required only for a few $j > \oal$.

In general, this way of updating $(v_j)$ for Dirichlet driven exchangeable stick-breaking processes is actually an adaptation of Algorithm 2 in \cite{Nea:00}, however, other marginal methods such as Algorithm 8 can also be exploited. In fact, by taking into account the underlying Dirichlet process, $P'$, of $(v_j)$, even a version of the slice sampler or the ordered allocation sampler could have been used. To conclude, we mention that for the simulation study in Section \ref{sec:4} of the main document we fixed the hyperparameters $\theta' = 1$, and $a=b=1$ for both geometric and Dirichlet driven exchangeable stick-breaking models.



\section{Ordered allocation variables}\label{sec:supp_example_di}
Here we discuss the set $\D_i$ of admissible moves for the updating of the ordered allocation variable $d_i$. Some general rules for determining $\D_i$ can be envisioned: (i) if $d_i$ is different from any other $d_j$, that is $D_{d_i}=\{i\}$, then $d_i$ cannot change, unless $d_i=k_n$; and (ii) $\D_i\subset\{1,\ldots,k_{i-1}+1\}$, so for larger $i$, there are more possible admissible moves, in particular, $d_1=1$ cannot change. 
As for illustration, let $n = 5$ and say that before updating $(d_i)$, $d_1,\ldots,d_5$ are such that the blocks of the partition in the least element order are $D_1 = \{1,3\}$, $D_2 = \{2,4\}$ and $D_3 = \{5\}$. Clearly $d_1=1$ cannot change. As for $d_2$, the admissible moves are $\D_2 = \{1,2\}$ thus $d_2$ will be sampled from
\[
\pr(d_2 =1 \mid \mathrm{rest}) \propto \tp_1 g(y_2 \mid \tx_1), \quad \pr(d_2 = 2 \mid \text{rest}) \propto \tp_2 g(y_2 \mid \tx_2).
\]
Say that we sample $d_2 = 1$, so that now $D_1 = \{1,2,3\}$, $D_2 = \{4\}$, $D_3 = \{5\}$. Since the blocks must be in least element order, the admissible moves for $d_3$ are $\D_3 = \{1,2\}$, hence $d_3$ will be sampled from
\[
\pr(d_3 =1 \mid \mathrm{rest}) \propto \tp_1 g(y_3 \mid \tx_1), \quad \pr(d_3 = 2 \mid \text{rest}) \propto \tp_2 g(y_3 \mid \tx_2).
\]
Assume we sample $d_3 = 1$ so now $D_1 = \{1,2,3\}$, $D_2 = \{4\}$,  and $D_3 = \{5\}$. Given that $D_2$ can not be an empty set, under the current configuration, the only admissible move for $d_4$ is $\D_4=\{4\}$, i.e. $d_4$ cannot change. Finally, the admissible moves for $d_5$ are $\D_5 = \{1,2,3\}$, and we will sample $d_5$ from
\[
\pr(d_5 =1 \mid \text{rest}) \propto \tp_1 g(y_5 \mid \tx_1), \quad \pr(d_5 = 2 \mid \text{rest}) \propto \tp_2 g(y_5 \mid \tx_2),
\]
\[
\pr(d_5 =3 \mid \text{rest}) \propto (1-\tp_1-\tp_2)g(y_5 \mid \tx_3).
\]
Finally, assuming that we sample $d_5 = 2$, the initial configuration $D_1 = \{1,3\}$, $D_2 = \{2,4\}$ and $D_3 = \{5\}$, is updated to $D_1 = \{1,2,3\}$ and $D_2 = \{4,5\}$.

In Section \ref{sec:acc_steps} we discuss an acceleration step that consists in randomly permuting the data points at each iteration of the sampler. Next we provide an example on how to modify the ordered allocation variables so to preserve the induced clustering structure, as well as the least element order of the partition induced by the modified variables. Let us consider again, as starting values of $(d_i)$ the ones corresponding to the partition $D_1 = \{1,3\}$, $D_2 = \{2,4\}$ and $D_3 = \{5\}$, so 
  $$(d_i) = (1,2,1,2,3).$$ 
Also consider the permutation $\rho = (2,1,3,5,4)$, i.e. $\rho(1) = 2, \rho(2) = 1, \ldots \rho(5) = 4$, so that the permuted data set is $(y_i') = (y_{\rho(i)}) = (y_2,y_1,y_3,y_5,y_4)$. The ordered allocation variables, $(d_i)$, induce the following clustering of data points:
$$\{y_1,y_3\} = \{y'_2,y'_3\},\quad\{y_2,y_4\} = \{y'_1,y'_5\},\quad\{y_5\} = \{y'_4\}.$$
Thus, the original partition $(\{1,3\},\{2,4\},\{5\})$ becomes $(D'_1,D'_2,D'_3) =(\{1,5\},\{2,3\},\{4\})$ with respect to the new labeling of the observations. 
Let us check that the ordered allocation variables, $(d'_i) = (1,2,2,3,1)$, that correspond to $(D'_j)$, can be obtained as explained in Section \ref{sec:acc_steps}, that is $d'_i=j$ if and only if $d_{\rho(i)}$ equals the $j$th distinct value to appear in $(d_{\rho(i)})$. We have
  $$(d_{\rho(i)}) = (d_2,d_1,d_3,d_5,d_4) = (2,1,1,3,2)$$
so that the distinct values of $(d_{\rho(i)})$ in order of appearance are $2,1,3$. Then,
\begin{align*}
&d_{\rho(1)} = d_2 = 2 \text{ is the first distinct value to appear in } (d_{\rho(i)}) \text{ hence } d'_1 = 1,\\
&d_{\rho(2)} = d_1 = 1 \text{ is the second distinct value to appear in } (d_{\rho(i)}) \text{ hence } d'_2 = 2,\\
&d_{\rho(3)} = d_3 = 1 \text{ is the second distinct value to appear in } (d_{\rho(i)}) \text{ hence } d'_3 = 2,\\
&d_{\rho(4)} = d_5 = 3 \text{ is the third distinct value to appear in } (d_{\rho(i)}) \text{ hence } d'_4 = 3,\\
&d_{\rho(5)} = d_4 = 2 \text{ is the first distinct value to appear in } (d_{\rho(i)}) \text{ hence } d'_5 = 1.
\end{align*}
We conclude that 
  $$(d'_i) = (1,2,2,3,1),$$ 
which in fact yields the partition in least element order $D'_1 = \{1,5\}, D'_2 = \{2,3\}, D'_3 = \{4\}$.





\section{Extended simulation study for the DP model}\label{sec:supp_illust}

\begin{figure}
\centering
\includegraphics[width=1\textwidth]{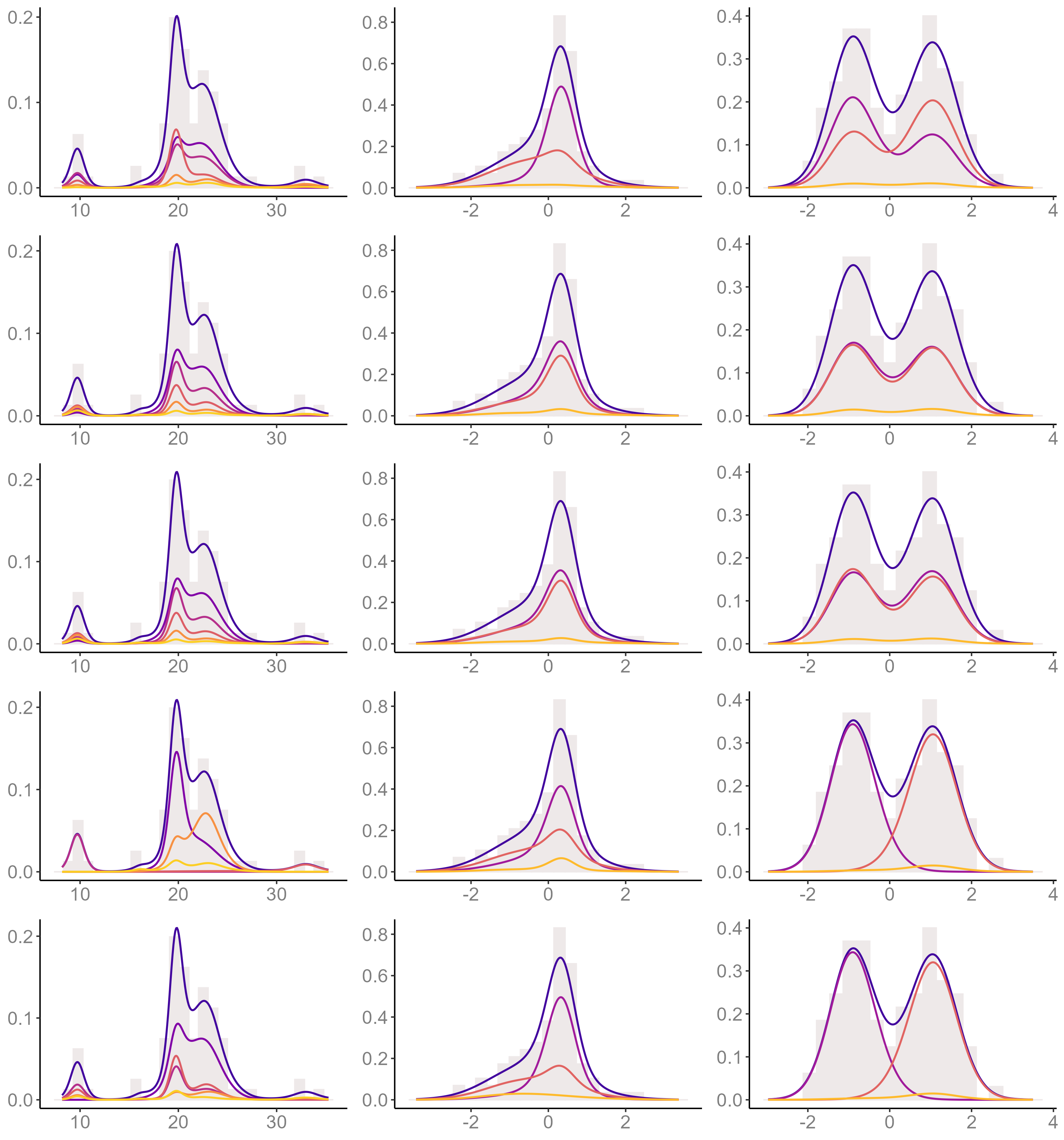}
\begin{small} 
\caption{Estimated density and weighted densities by component (colored lines) for the $\mathsf{galaxy}$ (left column), $\mathsf{leptokurtic}$ (middle column) and $\mathsf{bimodal}$ (right column) data sets, assuming a Dirichlet process mixing distribution. Implementation was made using a Marginal sampler (1st row), the ordered allocation samplers in Sections \ref{sec:OASsb} and \ref{sec:OAS} (2nd and 3rd row, respectively) including data permutations, the ordered allocation sampler in Section \ref{sec:OASsb} without data permutations (4th row) and the slice-efficient sampler (5th row).
}\label{fig:comp_lab}
\end{small}
\end{figure}

In this section we provide further illustrations of the two versions of the ordered allocation sampler derived in Sections \ref{sec:OASsb} and \ref{sec:OAS} of the main paper, and of the importance of the acceleration step in Section 3.3. We consider the Dirichlet process mixing prior and we repeat the simulation study of Section \ref{sec:4} implementing, together with Algorithm 8 in \cite{Nea:00} (Marginal), the dependent slice-efficient sampler by \cite{Kall:etal:11} (Conditional) and the sampler of Section \ref{sec:OASsb} with the acceleration step (OAS1), the sampler of Section \ref{sec:OAS} with the acceleration step (OAS2) and the sampler of Section \ref{sec:OASsb} without the acceleration step (OAS1*).
Table \ref{tab:OAS*} reports the IAT of the deviance ($D_v$) and number occupied components ($k_n$) for these 5 different samplers. 
In Figure \ref{fig:comp_lab} we show the estimated weighted densities, $\widehat{Q}_j$, of each component $j \leq K$, as well as the estimated density, $\widehat{Q} = \sum_{j=1}^K \widehat{Q}_j$, where $K$ is the largest index $j$, for which $\widehat{Q}_j$ is not identical to zero. We have computed 
\[
\widehat{Q}_j = \frac{1}{T}\sum_{t=1}^{T}\ \frac{n_j^{(t)}}{n} g\big(\cdot\,\big|\, x_j^{(t)}\big),
\]
for a window of $T = 10^{4}$ iterations after the burn-in period has elapsed. Here $x_1^{(t)},x_2^{(t)},\ldots$ are the sampled component parameters as labelled (or indexed) at iteration $t$, and $n_j^{(t)}$ is the number of data points assigned to the $j$th component at iteration $t$. In particular, for the ordered allocation samplers, components are labelled in the order in which they were discovered by the (possibly permuted) dataset at each iteration. For the marginal sampler, components are labelled this way at the first iteration, and at subsequent iterations relabelling only occurred to delete gaps, i.e. so that the first $k_n$ indexes, $j$, refer to the observed components. As for the slice sampler, components were never relabelled.

We will first focus on exploring the effects of the acceleration step in Section \ref{sec:acc_steps}. As can be observed in Figure \ref{fig:comp_lab} the ordered allocation sampler without data permutations (OAS1*, 4th row) is more prone to label components consistently throughout iterations, this is reflected through the propensity of $\widehat{Q}_j$ to be unimodal. In contrast, for the ordered allocation sampler that includes the acceleration step of Section \ref{sec:acc_steps} (OAS1, 2nd row) $\widehat{Q}_j$ has the shape of the estimated density $\widehat{Q}$. Thus $\widehat{Q}_j$ is multimodal when $\widehat{Q}$ is (as is the case of the $\mathsf{leptokurtic}$ dataset). This indicates that more label-switches occur in the OAS1, which is an expected consequence of the fact that by permuting data points, components are discovered in a distinct order. Comparing against the graphs of the marginal sampler (1st row), we see that by including the acceleration step, label-switches occur in a more similar way as they naturally do in marginal samplers, which is the only type of sampler where labels are completely irrelevant. In terms of algorithmic performance, in Table \ref{tab:OAS*} we see that the inclusion of data permutations at each iteration represents a significant improvement of the mixing properties, as the IAT corresponding to the OAS1 are much smaller than those of the OAS1*. The one exception is the IAT of $D_v$ for the $\mathsf{leptokurtic}$ dataset, which is very similar for the OAS1 and the OAS1*. This is due to the fact that, although this dataset comes from more than one Gaussian component, it only has one mode. Thus, it is not clear from which component does each data point come from, this is turn leads to frequent label-switches even if one does not permute the dataset. 

We now turn to explore the algorithmic performance of the OAS2 compared against that the OAS1, when the latter applies, i.e. the distribution of the size-biased permuted weights, $(\tp_j)$ is available. In Table \ref{tab:OAS*} we see that for the $\mathsf{galaxy}$ dataset, the IAT values of the OAS2 compare very well with those of the OAS1. For the other two datasets instead there is a difference between the IAT values of the OAS1 and the OAS2. To explain why this happens, recall that the key distinction between the OAS1 and the OAS2 is that the first one updates $(\tp_j)$ directly, while the OAS2  relies on the 
the indexes $(\alpha_j)$. In particular for the DP model, the OAS2 ignores the fact that the distribution of $(\tp_j)$ is available. As mentioned in the main paper, to update $(\alpha_j)$ the OAS2 first updates $(\alpha_j)_{j \leq k_n}$ (i.e. those of weights of occupied components) and later $(\alpha_j)_{j > k_n}$, hence swaps between $\alpha_j$ and $\alpha_i$, with $j \leq k_n < i$, mainly occur when occupied components are created or removed as a consequence of an update of  $(d_i)$.
Now, the $\mathsf{leptokurtic}$ and $\mathsf{bimodal}$ datasets were both simulated from a mixture of two (more or less) balanced Gaussian components. Since we have implemented a Gaussian mixture, at most iterations the sampler will effectively recognize that there are only two large occupied components (see the 2nd and 3rd columns of Figure \ref{fig:comp_lab} for an illustration). Thus the mixing of $(\alpha_j)$ will be affected because components are rarely created of removed. Furthermore, 
as typically there will be only two large \comillas{occupied} weights and the rest of them will be very small, 
it is extremely unlikely that their indexes get swapped. 
Instead, the $\mathsf{galaxy}$ dataset was not generated from a Gaussian mixture, which forces the model to rely on different Gaussian components of varying sizes to estimate the density (cf. 1st column of Figure \ref{fig:comp_lab}). This means that the number of occupied components will change frequently and some of the \comillas{occupied} weights will be small at many iterations thus facilitating the mixing of weights' indexes. 


\begin{table}[htbp]
\centering
\begin{tabular}{c | c | c c c c c}
\hline
\multicolumn{2}{c}{} & Marginal & OAS1 & OAS2 & OAS1* & Conditional \\ \hline
\multirow{ 2}{*}{$\mathsf{galaxy}$} & $D_v$ & 12.30(0.23) & 23.76(0.57) & 26.76(0.84) & 106.9(7.98) & 119.2(9.95) \\
& $k_n$ & 13.68(0.25) & 32.49(0.81) & 36.36(1.09) & 115.2(7.18) & 190.2(16.3)  \\ \hline
\multirow{ 2}{*}{$\mathsf{leptokurtic}$} & $D_v$ & 22.42(0.54) & 25.81(0.63) & 35.98(1.00) & 26.92(0.75) & 120.5(10.3) \\
& $k_n$ & 9.26(0.13) & 18.99(0.41) & 27.97(0.88) & 44.48(0.60) & 100.8(7.18)  \\ \hline
\multirow{ 2}{*}{$\mathsf{bimodal}$} & $D_v$ & 7.84(0.16) & 13.87(0.35) & 20.82(0.57) & 50.47(3.20) & 35.61(1.68)  \\ 
& $k_n$ & 6.30(0.07) & 13.38(0.22) & 25.43(0.87) & 39.28(1.33) & 52.00(2.18)  \\ \hline
\end{tabular}
\caption{IAT of $D_v$ and $k_n$ (standard errors are shown in parenthesis) for Algorithm 8 in \cite{Nea:00} (Marginal), the ordered allocation samplers in Sections \ref{sec:OASsb} and \ref{sec:OAS} with data permutation (OAS1 and OAS2, respectively), the ordered allocation sampler in Section \ref{sec:OASsb} without data permutations (OAS1*) and the dependent slice-efficient sampler (Conditional), obtained by fitting a DP model to the $\mathsf{galaxy}$, $\mathsf{leptokurtic}$ and $\mathsf{bimodal}$ datasets.\label{tab:OAS*}}
\end{table}


\end{document}